\documentclass{pasa}%

\usepackage{graphicx}
\usepackage{times}
\usepackage{url}
\usepackage{epsfig}
\usepackage[flushleft]{threeparttable}
\usepackage{pdfpages}
\usepackage{array}
\usepackage{comment}
\usepackage{amsmath}	
\usepackage{amssymb}	
\usepackage{aas_macros}
\usepackage{hyperref} 
\hypersetup{colorlinks,citecolor=blue,linkcolor=blue,urlcolor=blue}

\newcommand{\refbf}{}

\title[SkyMapper AlertSDP: Alert Science Data Pipeline]{SkyMapper Optical Follow-up of Gravitational Wave Triggers: Alert Science Data Pipeline and LIGO/Virgo O3 Run} 

\author[S.-W. Chang et al.]
{Seo-Won Chang$^{1,2,3,4,5}$\thanks{E-mail: seowon.chang@anu.edu.au or seowon.chang@snu.ac.kr},
Christopher A. Onken$^{1,3}$, 
Christian Wolf$^{1,2,3}$,
Lance Luvaul$^{1}$,
Anais M\"oller$^{1,6}$,
Richard Scalzo$^{1,7}$,
Brian P. Schmidt$^{1}$,
Susan M. Scott$^{2,3,8}$,
Nikunj Sura$^{1}$,  
and Fang Yuan$^{1}$

\affil{$^{1}$Research School of Astronomy and Astrophysics, The Australian National University, Canberra, ACT 2611, Australia}
\affil{$^{2}$ARC Centre of Excellence for Gravitational Wave Discovery (OzGrav), Australia}
\affil{$^{3}$Centre for Gravitational Astrophysics, The Australian National University, ACT 2601, Australia}
\affil{$^{4}$SNU Astronomy Research Center, Seoul National University, 1 Gwanak-rho, Gwanak-gu, Seoul 08826, Korea}
\affil{$^{5}$Astronomy program, Dept. of Physics \& Astronomy, Seoul National University, 1 Gwanak-rho, Gwanak-gu, Seoul 08826, Korea}
\affil{$^{6}$Universit{\'e} Clermont Auvergne, CNRS/IN2P3, LPC, F-63000 Clermont-Ferrand, France}
\affil{$^{7}$Centre for Translational Data Science, University of Sydney, Darlington, NSW 2008, Australia}
\affil{$^{8}$Research School of Physics, The Australian National University, Canberra, ACT 2601, Australia}
}

\jid{PASA}
\doi{10.1017/pas.\the\year.xxx}
\jyear{\the\year}

\hypersetup{draft}

\begin{document}

\begin{frontmatter}
\maketitle

\begin{abstract}
We present an overview of the SkyMapper optical follow-up program for gravitational-wave event triggers from the LIGO/Virgo observatories, which aims at identifying early GW170817-like kilonovae out to $\sim 200$\,Mpc distance. We describe our robotic facility for rapid transient follow-up, which can target most of the sky at $\delta<+10\deg $ to a depth of $i_\mathrm{AB}\approx 20$\,mag. We have implemented a new software pipeline to receive LIGO/Virgo alerts, schedule observations and examine the incoming real-time data stream for transient candidates. We adopt a real-bogus classifier using ensemble-based machine learning techniques, attaining high completeness ($\sim$98\%) and purity ($\sim$91\%) over our whole magnitude range. Applying further filtering to remove common image artefacts and known sources of transients, such as asteroids and variable stars, reduces the number of candidates by a factor of more than 10. We demonstrate the system performance with {\refbf data} obtained for GW190425, a binary neutron star merger detected during the LIGO/Virgo O3 observing campaign. In time for the LIGO/Virgo O4 run, we will have deeper reference images allowing transient detection to $i_\mathrm{AB}\approx $21\,mag.
\end{abstract}


\end{frontmatter}

\section{Introduction}
\label{sec:introduction}
The observable signature of merging binary neutron stars (BNS) was revealed for the first time by the event GW170817, which started with a gravitational-wave (GW) chirp signal detected by the Advanced LIGO/Virgo detectors and was then followed by electromagnetic (EM) observations over a broad range of wavelengths from $\gamma$-rays to radio \citep{PhysRevLett.119.161101,Abbott2017ApJ...848L..12A}. One of the most promising outcomes is luminous EM emission in the optical and near-infrared bands after the final coalescence of the stars (e.g., \citealt{Andreoni2017PASA...34...69A,Arcavi2017Natur.551...64A,Coulter2017Sci...358.1556C,Cowperthwaite2017ApJ...848L..17C,Drout2017Sci...358.1570D,Kilpatrick2017Sci...358.1583K,Kasliwal2017Sci...358.1559K,Smartt2017Natur.551...75S}). This fast-evolving transient, referred to as a kilonova, was predicted to be powered by the radioactive decay of heavy elements via rapid neutron capture processes \citep{Li1998ApJ...507L..59L,Metzger2010MNRAS.406.2650M,Barnes2013ApJ...775...18B,Tanaka2013ApJ...775..113T,Fernandez2016ARNPS..66...23F}. We now know that rapid EM follow-up (on the timescale of hours) provides a wealth of information on the nature of the progenitor {\refbf \citep{2017ApJ...850L..40A,2018MNRAS.481.1908K}}, its environment {\refbf \citep{Levan2017ApJ...848L..28L,2020ApJ...905...21A}} and explosion mechanism {\refbf \citep{2017Natur.551...71T,2018Natur.554..207M}}. In particular, the spectral evolution of such events is key to understanding the merger process and the origin of rare heavy elements {\refbf \citep{2017Natur.551...80K,2019MNRAS.tmpL..14K,2019PhRvL.122f2701W}}. Furthermore, the identification of an EM counterpart is needed to pinpoint the host galaxy, thus determining the redshift {\refbf \citep{Coulter2017Sci...358.1556C,2017ApJ...848L..31H}}. 

Photometric and spectroscopic data of kilonovae provide a novel probe into the physics and nature of the BNS themselves (e.g., NS radius or mass ratio) and into the process and end-product of mergers. {\refbf Although GW170817 remains the only kilonova that was both spectroscopically confirmed and associated with a GW signal}, it allows us to better understand how different ejecta components (with different lanthanide fractions) contribute its EM emission from early to late times after the merger. However, little is known about the earliest stages of the kilonova, because the EM coverage of GW170817 started more than 10 hours after the event. Hence, we have no constraints yet on possible emission from the outermost layers of the ejecta that may have faded after a few hours. For example, \citet{Metzger2015MNRAS.446.1115M} proposed a candidate precursor of kilonova emission, caused by $\beta$-decay of free neutrons in the outermost ejecta, which can increase the luminosity of the EM source by over an order of magnitude during the first hour after the merger. \citet{Metzger2019LRR....23....1M} mentions $r$-process heating or radioactive decay of free neutrons. And further mechanisms such as jet/wind re-heating could play a similar role in producing enhanced luminosity at early times (e.g., \citealt{Metzger2018ApJ...856..101M}). Only high-cadence observations in the first few hours after the merger can test these predictions in detail and reveal the source of the blue emission \citep{2018ApJ...855L..23A}.

The SkyMapper optical wide-field telescope in Australia (see Section \ref{subsec:SkyMapper Telescope}) is one of the facilities that can discover early GW170817-like kilonovae and is probably the pivotal optical facility for events that occur in the Southern sky between the end of the Chilean night and late in the Australian night. 
The size of the GW localisation area \citep[a few hundred deg$^2$, see][]{2020LRR....23....3A} is {\refbf usually much larger than the field of view (FoV) of our camera} ($5.7$ deg$^2$). 
{\refbf While the GW triggers in the third LIGO/Virgo observing run (O3) had a median localisation area of 4480 deg$^2$ \citep{2020ApJ...905..145K}, the area shrinks quickly for events with stronger signals.}
{\refbf A synoptic tiling strategy covering the high-probability sky area for the counterpart will usually best exploit our wide FoV, and our tiles follow the basic tiling scheme of the general-purpose SkyMapper Southern Survey \citep[see Section~\ref{subsec:SkyMapper Telescope}]{Onken2019}. In contrast, an alternative galaxy-targeted approach makes more sense for telescopes with smaller FoV and for nearby events such as GW170817, where it has proven to be effective. However, for events at distance larger than 50~Mpc, which are expected to be by far the most common, our FoV contains usually several possible host galaxies and empty tiles will be rare. }
Crucially, we can detect transient candidates in real-time as we have reference images for subtraction over the full hemisphere ({\refbf at least in some passbands}, see Section \ref{subsubsec:reference selection}), which are deep enough to detect GW170817-like kilonovae at distances up to 200\,Mpc as well as the rising part of kilonova light curves in more nearby cases.


In the O3 run, 56 gravitational-wave events from compact binary systems were detected, which is five times more than reported during the first two observing runs. {\refbf Only two of them had significant ($>85\%$) initial probability of being BNS events: S190425z \citep{2019GCN.24168....1L} and S190901ap \citep{2019GCN.25606....1L}. However, after re-analysis with additional background statistics the latter is no longer considered a significant candidate \citep{2020arXiv201014527A}.} S190425z, a.k.a. GW190425 and known in Australia as the ANZAC Day event, was also confirmed as the second case of gravitational waves from a binary neutron star inspiral \citep{2020ApJ...892L...3A}. The system is noteworthy for a total mass of 3.4$M_{\odot}$, which exceeds that of known Galactic BNS and may suggest that not all binary neutron stars are formed in the same way (\citealt{Romero-Shaw2020arXiv200106492R, Safarzadeh2020arXiv200104502S}). No EM counterpart was found by SkyMapper or any other facility, because (i) the sky localisation of this event was poorly constrained with a 90\% confidence area of 8\,284 deg$^{2}$, (ii) it is expected to be much fainter than GW170817 in the optical due to its distance, {\refbf and (iii) it may be intrinsically faint due to a low ejecta mass or unfavourable viewing angle.} In preparation for the next LIGO/Virgo observing run (O4), we describe here our facility and its performance with the current processing pipeline.

For an autonomous selection of transient candidates, it is key to maximise the recovery rate and minimise the false-positive rate; in order to reduce the volume of human intervention required to identify the likely source of interest, on which detailed follow-up observations may be triggered. This process faces two challenges: (1) Image artefacts appear in the subtraction process and pose as transient candidates; this is often addressed with machine learning approaches that separate astrophysical sources from spurious detections in a difference image (e.g., \citealt{Bailey2007ApJ...665.1246B, Bloom2012PASP..124.1175B, Brink2013MNRAS.435.1047B, Wright2015MNRAS.449..451W, Goldstein2015AJ....150...82G,Duev2019MNRAS.489.3582D}).
(2) Astrophysical {\refbf foreground and background} transients produce a fog of events that are not related to the GW event, and these are usually too numerous for simultaneous follow up. In this work, we use an ensemble-based transient classifier to reject spurious sources and refer to catalogues of known sources to label other types of variables. 

In this paper, we present an overview of the SkyMapper follow-up program of GW triggers. In Section \ref{sec:facility and process flow}, we describe our optical facility and the AlertSDP pipeline in detail, from observing strategy to real-time data processing and transient identification. In Section \ref{sec:New RB Classifier}, we introduce an ensemble-based machine learning approach for real-bogus classification. We also present metrics for evaluating the performance of the classifier. In Section \ref{sec:application of new metric}, we present a real case and discuss the resulting transient statistics. In Section \ref{sec:summary}, we close with an outlook to future work and the LIGO/Virgo O4 run. Throughout the paper we use the AB magnitude system.

\section{EM counterpart searches with the SkyMapper facility}
\label{sec:facility and process flow}

\subsection{SkyMapper Telescope}
\label{subsec:SkyMapper Telescope}
SkyMapper is a 1.35m modified-Cassegrain telescope located at Siding Spring Observatory (SSO) in New South Wales, Australia \citep{Wolf2018PASA...35...10W}. The telescope has a wide field-of-view of 2.34 $\times$ 2.37 deg$^2$, a \(uvgriz\) filter set  \citep{Bessell2011PASP..123..789B} and a mosaic of 32 2k$\times$4k CCD detectors with a pixel scale of 0.5 arcsec/pixel. It is owned and operated by the Australian National University (ANU) and, most importantly for EM follow-up, it is a robotic facility.

\begin{figure*}[!t]
\includegraphics[width=\linewidth]{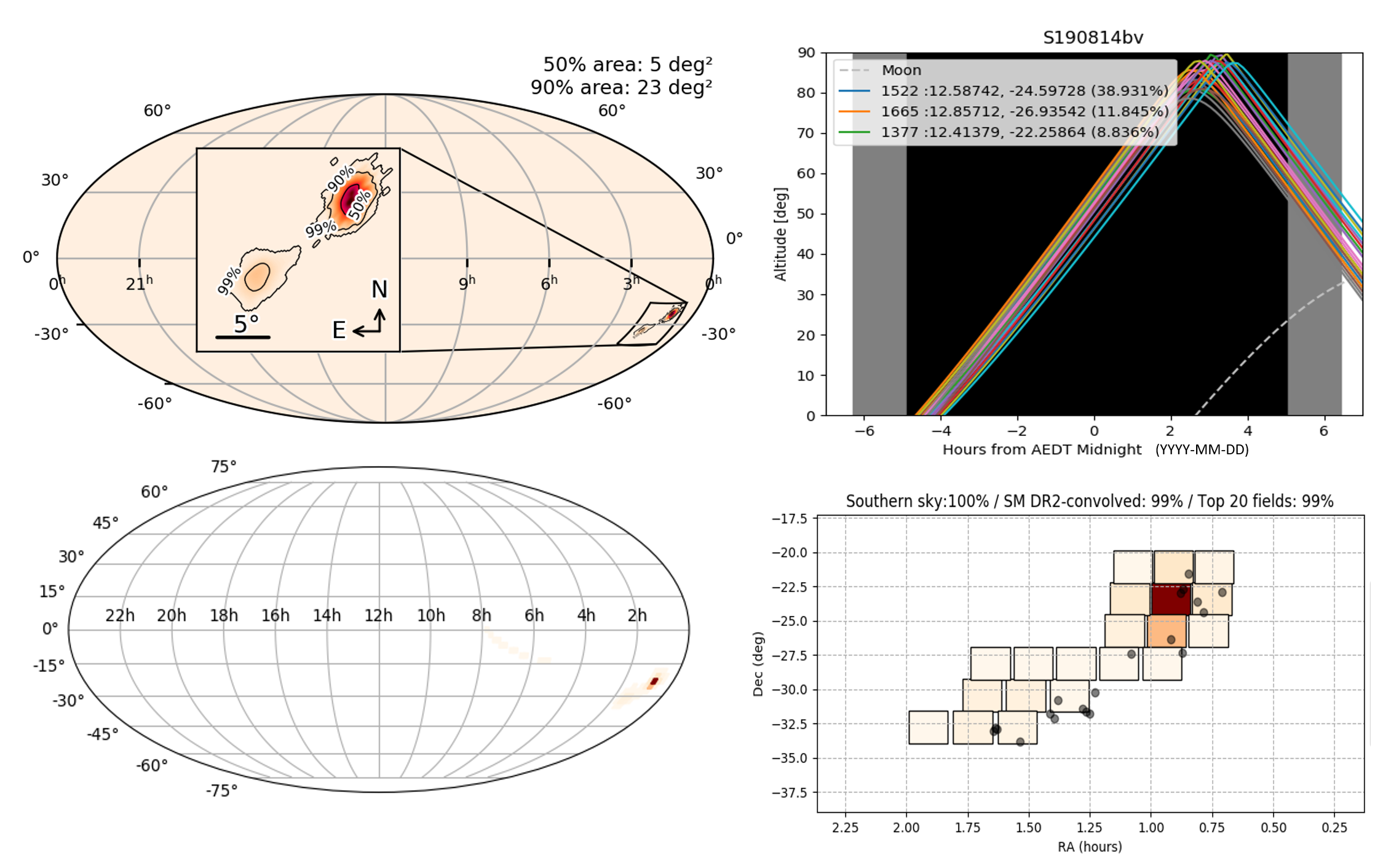}
\caption{Top left: LIGO/Virgo probability sky map {\refbf for S190814bv produced with the {\it ligo.skymap} module \citep{2016ApJS..226...10S}. The inset shows the most probable area for the optical counterpart. Darker colours correspond to higher-probability sky regions, and contours enclosing regions with 50\%, 90\%, and 99\% of the probability are indicated.} Bottom left: probability map convolved with the coverage of reference images in SkyMapper DR2. Bottom right: zoomed-in map of the 20 highest-probability fields selected for the search; here, one field alone has $\sim 39$\% probability of containing the GW source. Symbols are bright galaxies from the 2MASS redshift survey. Top right: Observability plot for the top 20 fields with telescope altitude, night-time range and Moon separation. }
\label{figure: SkyMapper probability skymap and target observability}
\end{figure*} 

The main purpose of the telescope is the SkyMapper Southern Survey (SMSS), a hemispheric sky atlas which has been underway since 2014 (DR1: \citealt{Wolf2018PASA...35...10W}; DR2: \citealt{Onken2019}). For point sources with SNR$>$5, the expected photometric depth of single-epoch 100-sec images is $u$=19.5, $v$=19.5, $g$=21, $r$=20.5, $i$=20, and $z$=19 mag. Alongside, SkyMapper has been used to {\refbf search for} extragalactic transients, including low-redshift Type Ia supernovae \citep{Scalzo2017PASA...34...30S, 2019IAUS..339....3M} {\refbf as well as optical counterparts to} GW events (GW170817: \citealt{Abbott2017ApJ...848L..12A, Andreoni2017PASA...34...69A}) and fast radio bursts (e.g., \citealt{2015MNRAS.447..246P, 2018MNRAS.478.1209F,2019MNRAS.488.2989F, 2019MNRAS.486.3636P, 2019ATel13008....1C}). Different needs {\refbf made the} SMSS and the transient searches develop {\refbf separate} data reduction software: the Science Data Pipeline (SDP: \citealt{Luvaul2017ASPC..512..393L, Wolf2018PASA...35...10W}) {\refbf for the SMSS and the Subtraction} Pipeline (SUBPIPE; \citealt{Scalzo2017PASA...34...30S}) {\refbf for the Transient Survey}. SUBPIPE was used for EM data analysis before LIGO/Virgo ER14 (Engineering Run 14), but for the O3 run we updated it to an automated real-time pipeline (AlertSDP: Alert Science Data Pipeline) that borrows many features from the SDP (see Section \ref{subsec:alert response} and \ref{subsec:AlertSDP} for details).

\subsection{Alert Response}
\label{subsec:alert response}
We continuously listen to the live stream of LIGO/Virgo public alerts for compact binary merger candidates\footnote{\url{https://emfollow.docs.ligo.org/userguide/}}. The stream is distributed through the Gamma-ray Coordinates Network (GCN) using the {\it pygcn} {\sc Python} module\footnote{\url{https://github.com/lpsinger/pygcn}}. We developed a robotic alert handler that extracts relevant information, ingests the GW event into our database, downloads the HEALPix 3D localisation map (skymap), prioritises areas for follow-up, and generates a list of observations for SkyMapper. 

A first preliminary GCN notice is automatically issued for a superevent within 1--10 minutes after the GW trigger. From this notice, our robotic handler initiates rapid-response {\it search} observations by convolving the probability in the BAYESTAR skymap \citep{PhysRevD.93.024013} with the tile pattern of the SkyMapper Southern Survey \citep{Onken2019}, ranking the fields by probability integrated per-field and then selecting the top 20 fields with existing $i$-band reference images ($\sim$100 $\mathrm{deg}^2$ total area). Figure \ref{figure: SkyMapper probability skymap and target observability} shows the sky maps for a real alert and an observability map of the selected target fields.

\begin{figure}[!t]  
\centering
\includegraphics[width=\linewidth]{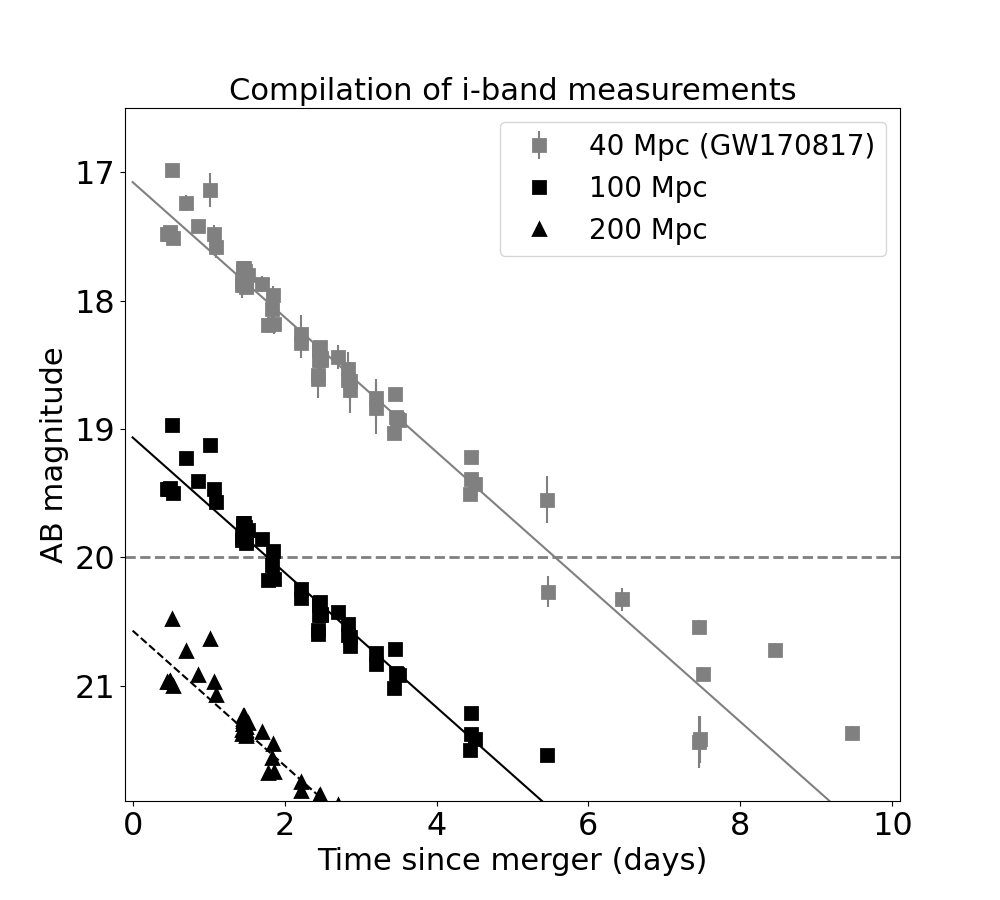}
\caption{The $i$-band light curve for the GW170817 kilonova, at different distances: at the true distance of 40 Mpc (top), shifted to 100 Mpc (middle) and 200 Mpc (bottom); solid lines are power-law decay fits. The dashed line at $i_\mathrm{AB}=20$ marks our typical 5-$\sigma$ magnitude limit in 100\,sec exposures (data were taken from various literature sources; AST3-2: \citealt{2017SciBu..62.1433H}; B\&C: \citealt{Utsumi2017PASJ...69..101U}; DECam: \citealt{Cowperthwaite2017ApJ...848L..17C}; Gemini: \citealt{Kasliwal2017Sci...358.1559K}; LaSilla: \citealt{Smartt2017Natur.551...75S}; LCO: \citealt{Arcavi2017Natur.551...64A}; Magellan: \citealt{Shappee2017Sci...358.1574S}; Pan-STARRS: \citealt{Smartt2017Natur.551...75S}; REM: \citealt{Pian2017Natur.551...67P};  SkyMapper: \citealt{Andreoni2017PASA...34...69A}; Swope: \citealt{Drout2017Sci...358.1570D};T80S: \citealt{Diaz2017ApJ...848L..29D}; VLT: \citealt{Tanvir2017ApJ...848L..27T}; VST: \citealt{Pian2017Natur.551...67P}). }\label{figure: GW170817 kilonova lightcurve}
\end{figure}

In this search stage, we obtain two $100\,\mathrm{s}$ exposures in $i$-band for each field, separated by $\sim 8$\,minutes. Requiring a twin detection in the two images eliminates moving objects from the candidate list. 
{\refbf The typical rate of motion for main-belt asteroids, 0.625 arcsec/min
near opposition \citep{1996AJ....111..970J}, provides a sufficient $\sim 5$ arcsec spacing between observations.} With a depth of $i\approx$ 20\,mag, we can identify kilonovae that are $\sim 3$\,mag (16$\times$) fainter than the kilonova of GW170817 was ten hours after the GW trigger. Hence, we could detect kilonovae for GW170817-like events out to 4$\times$ the distance of GW170817, or 160\,Mpc. Figure \ref{figure: GW170817 kilonova lightcurve} shows the lightcurve of the GW170817 kilonova shifted to different distances up to 200 Mpc. The declining nature of the $i$-band light curve suggests that kilonovae may be more luminous during the first ten hours. We thus assume that we might be able to detect early kilonova emission, just hours after a BNS merger, out to distances of 200\,Mpc or beyond.

{\refbf We search in $i$-band for three reasons: (i) kilonovae are expected to be red due to the large opacity of neutron-rich ejecta, although at early times (within an hour of the merger) blue emission from hotter, polar ejecta may be expected (e.g., \citealt{2018NatCo...9.4089T})}, (ii) the $i$-band covers the largest sky area with deep reference images in the SMSS that can be used for real-time image subtraction, and therefore provides the greatest search volume for kilonovae, and (iii) the dominant source of transient events, flares on M\,stars too faint to be seen in quiescence, is nearly irrelevant in $i$-band \citep{Chang2020MNRAS.491...39C}. The aim of the search phase is to find a counterpart as fast as possible, report it to facilities around the world and get SkyMapper itself into monitoring mode. We expect that the search phase provides a full set of possible counterparts within four hours (limited by image processing) after the GW trigger, if the trigger occurs during darkness and the sky is clear. 
Later, possibly within 4 hours for BNS or NSBH events, an updated LALInference skymap \citep{PhysRevD.91.042003} will be distributed, including an updated sky localisation and source classification. This sometimes leads to a significant change in the sky localisation, and the SkyMapper observing plan may be modified and re-executed as a result.

At any time, once a position of a likely kilonova transient is identified, we can manually switch from the search phase to a continuous high-cadence monitoring of the new source. Observations of an early kilonova with a cadence of 2\,min would reveal structure in the lightcurve arising from shocks induced by the kilonova ejecta (e.g., \citealt{Metzger2019LRR....23....1M}). We will also get high-cadence, multi-band light curves for several consecutive nights after any kilonova discovery. The monitoring strategy will change depending on the colour, luminosity, and fading time-scale of the kilonova candidates. If the optical counterpart gets identified by other groups before SkyMapper can observe, we will do only light-curve monitoring, as was the case for GW170817 \citep{Andreoni2017PASA...34...69A}.

\begin{figure}[!t]
\centering
\includegraphics[width=\linewidth]{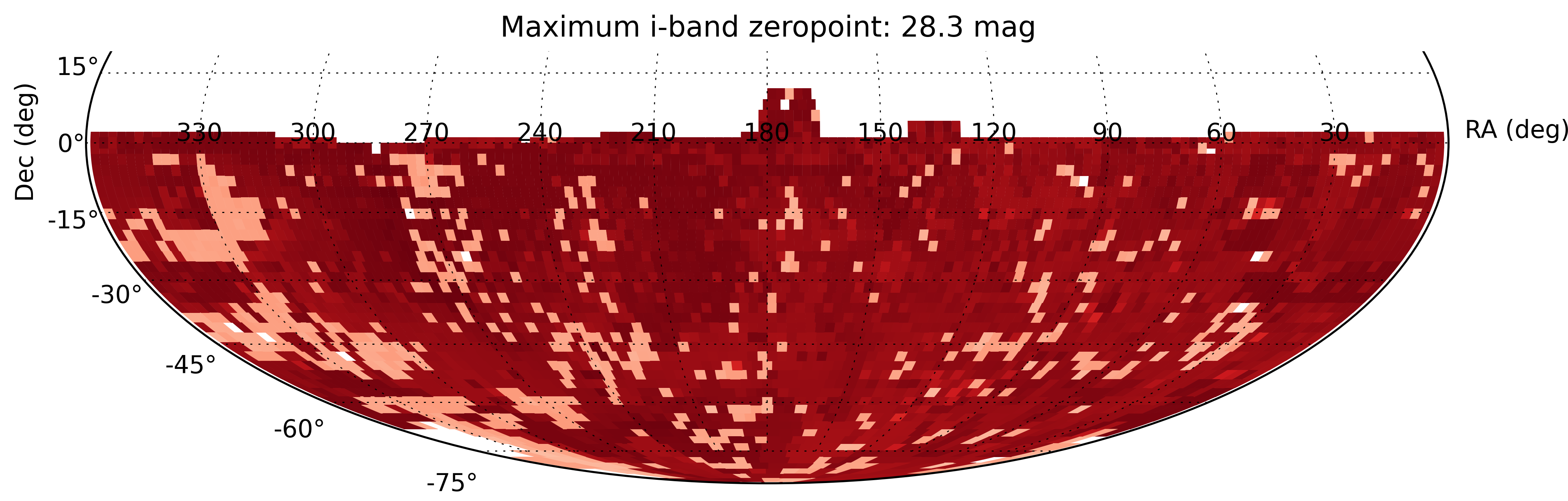}
\includegraphics[width=\linewidth]{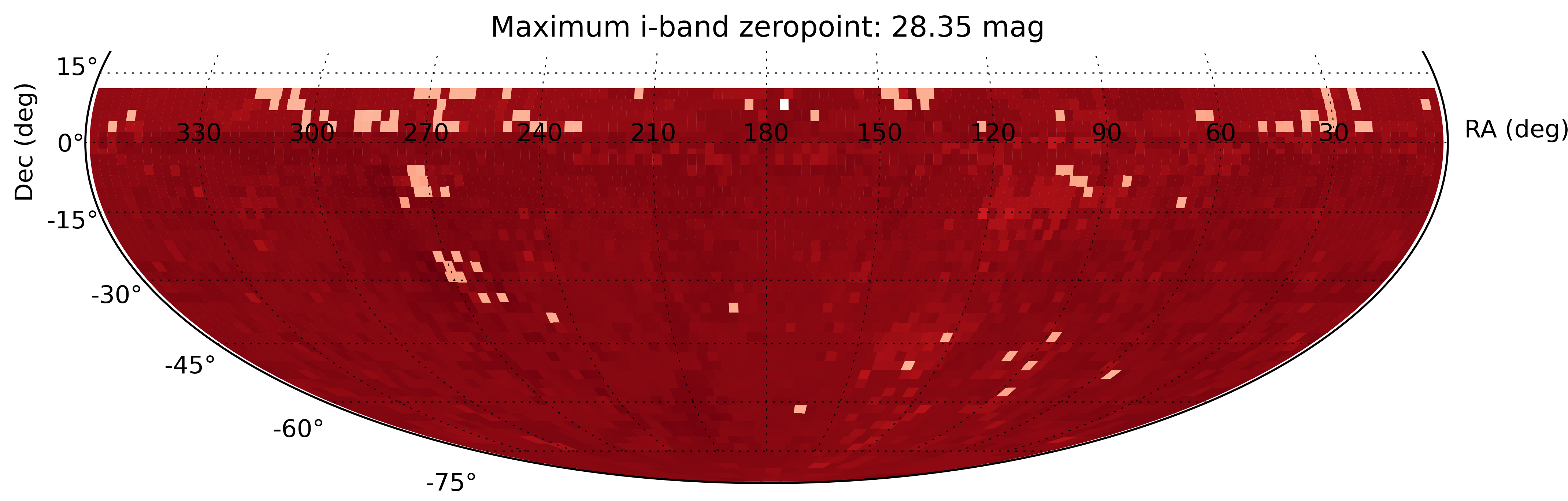}
\caption{SkyMapper \(i\) band coverage in DR2 (top) and DR3 (bottom). Darker colours resemble deeper reference images used for subtraction.}
\label{figure: SkyMapper i-band coverage}
\end{figure}

\subsection{AlertSDP: Alert Science Data Pipeline}
\label{subsec:AlertSDP}
The AlertSDP is a new pipeline created from two existing software packages, which are the SkyMapper Science Data Pipeline, SDP, and SUBPIPE, which was used for discovering low-redshift supernovae (see Section \ref{subsec:SkyMapper Telescope}). The SDP has provided the general process control framework as well as improved calibration and masking treatment, while the subtraction and transient classification have evolved from SUBPIPE components. First, each raw image is pre-processed and calibrated. After that, $i$-band reference images for image subtraction are selected, and transient candidates detected on difference images are uploaded into a database. The database also includes external catalogues to assist transient classification.

\subsubsection{Ingest and calibration of new images}
\label{subsubsec:pipeline flow}
Raw SkyMapper images from Siding Spring Observatory are transferred in real time via Ethernet to a 64-core server at Mount Stromlo Observatory. Data transfer takes 5--7 seconds per frame and can be done while the next exposure is already being taken. To activate the AlertSDP, the Linux {\it inotify} mechanism detects the arrival of an image from the telescope immediately and initiates the data processing.

The raw data are processed into scientific data products using typical SDP procedures, including bias correction, flat-fielding, defringing, bad-pixel masking, and generation of an astrometric solution (see \citealt{Wolf2018PASA...35...10W} and \citealt{Onken2019} for details). A significant structural change from the SDP to the AlertSDP is a different parallelisation paradigm designed to minimise the time between image acquisition and production of transient candidates: while the SDP processes images in parallel with all the component CCDs treated serially, the AlertSDP processes the 32 CCDs of each image in parallel and up to 4 images concurrently.

\subsubsection{Selection of reference images}
\label{subsubsec:reference selection}
Reference images are required for image subtraction and real-time transient searching. We take advantage of the nearly all-Southern-sky coverage ($>$98 \%) of the deeper Main Survey exposures from the SMSS in \(i\)-band ($\lambda_\mathrm{centre}$/$\Delta \lambda$ = 779/140 nm). Figure \ref{figure: SkyMapper i-band coverage} shows the area of sky covered by SkyMapper DR2 \citep{Onken2019} and by DR3, which was released {\refbf to the Australian community} in February 2020 in the late stages of the O3 run. DR3 {\refbf expands the deep coverage by including survey observations from March 2014 to October 2019}\footnote{\url{ http://skymapper.anu.edu.au/data-release/dr3/}}.

From any new image, we subtract each available reference image. The best possible reference image combines a small PSF and a large overlap with the new image. The reference images come with a range of point-spread functions (PSF), and because the SMSS obtains repeat images with dithering to cover the sky homogeneously, the best-available reference PSF changes discontinuously with sky location. Small overlaps lead to badly determined convolution kernels and thus bad subtractions. Hence, we require at least 15\% overlap area for mosaic frames and at least 5\% overlap for each individual CCD. Following a common strategy of other transient surveys, we require that the reference image should be taken at least two weeks prior to a given new epoch.

\subsubsection{Subtraction of images}
\label{subsubsec:DoSubtraction}
We perform image subtraction on each field of overlapping new/reference image pairs. First, we resample each of the reference mosaic frames onto the new image with {\sc SWarp} \citep{2002ASPC..281..228B}. Next, we convolve those pairs of images to a common PSF using {\sc HOTPANTS} \citep{Becker2015ascl.soft04004B}, which implemented the popular image subtraction algorithm by \citet{Alard1998ApJ...503..325A}. Solving for the convolution kernel is a crucial step to equalising the PSFs of the reference and new images. The position-dependent PSF variation in both images is modelled as a linear combination of basis functions, and by default we choose a 2D polynomial of order two. 
Next, the flux level of the subtracted image is normalised to that of the reference image. Since the photometric calibration is directly tied to the SMSS photometry, this approach has the advantage of allowing explicit calculation of zero-point corrected magnitudes. The preliminary calibrated magnitude and error used for lightcurve construction is:
\begin{align}
    m_{\rm sub} &= -2.5 \log f_{\rm sub} + {\rm ZP_{\rm ref}}, \notag \\
    \delta m_{\rm sub} &= \sqrt{(1.0857 \times \delta f_{\rm sub}/f_{\rm sub} )^{2} + {\rm \delta ZP_{\rm ref}}^{2}}, \notag
\end{align} where ${f_{\rm sub}}$ is the difference flux on the subtracted image and ${\rm ZP_{\rm ref}}$ is the zeropoint of the reference image. Here, we use the 15 arcsec (30-pixel-diameter) aperture as a total-magnitude reference. The zeropoint error $\delta {\rm ZP_{\rm ref}}$ contributes little, since only images with robust zeropoints are included in SMSS data releases.

Finally, we run SExtractor \citep{Bertin1996A&AS..117..393B} on the subtracted images to produce a list of transient candidates and their associated metadata, using a low detection threshold of 1.5$\sigma$ above the background. Gaussian filters in SExtractor are applied to an image prior to the detection of sources. Only sources with a signal-to-noise ratio below 2 are referred to as sub-threshold detections.

\subsection{Transient classification}
\label{subsec:DoClassify}
Until the end of O3, we classified transient candidates with a random forest (RF) classifier trained on earlier supernova survey data \citep{Scalzo2017PASA...34...30S}. We computed a set of features derived from difference images of individual candidates, similar to those proposed in \citet{Bloom2012PASP..124.1175B}. The main difference was that three additional checks were made on each candidate: (i) we removed any detections where the measurement of shape parameters (e.g., FWHM, elongation) was significantly larger than the median quantities for sources found in the new science images, and (ii) we matched each of them to the nearest object ($<$30 arcsec) seen on the reference frame, which will be associated with host galaxies or bright stars that might be poorly subtracted. Then, the RF model assigned a real-bogus score (RBscore) from 0 (artefacts = bogus) to 100 (transients = real) to all the detected candidates.

In this work, we added a new classifier (XGBoost) and new training sets to exploit our improved image processing. A combination of classifiers is often more accurate than a single classifier, and thus we introduce a new metric (Tscore) that combines RBscore and XGBoost score to compensate for the weakness of each classifier (see Section \ref{sec:New RB Classifier} for the full description and performance).

For each transient candidate, we identify {\refbf whether it may be a known variable or Milky Way source by cross-matching it against catalogues of stars, variable objects and quasars, using a 3 arcsec search radius. We also reject known solar system} objects with the SkyBoT cone-search service \citep{Berthier2006ASPC..351..367B}.  
Our pre-selected categories are as follows:
\begin{itemize}
    \item We define a "Star" sample using parallaxes and proper motions (PPM) from Gaia eDR3 \citep{2016A&A...595A...1G,2020arXiv201201533G}, requiring that the significance of a PPM signal\footnote{$S^2_{\rm PPM} = ((pmra/pmra\_error)^{2}+ (pmdec/pmdec\_error)^{2} + (\max(0,parallax)/parallax\_error)^{2} )/3$} is $S_{\rm PPM}>3$.
    
    \item For Gaia eDR3 objects with lower PPM\textunderscore SN or no PPM information, we use a "GaiaSource" label.
    
    \item We use the label "Var" for all sources cross-matched to the AAVSO International Variable Star Index (\citealt{Watson2006SASS...25...47W}; version 19 October 2020). It contains mainly known variable stars and a few QSOs.
    
    \item We assign a label of "Quasar" to sources matched to the Milliquas catalogue  (\citealt{Flesch2015PASA...32...10F}, version 7.0a). 
    
    
    \item For detections that are too close to very bright stars and carry a risk of being an optical reflection (defined by the algorithm in item 3 of Section~2.6.1 from \citealt{Onken2019}), we use a "BrightStar" label.
\end{itemize}
In addition, image mask data allows us to reject sources as being likely spurious, when they could be affected by bad pixels, neighbouring saturated pixels, or cosmic ray hits. Only candidates with Tscore greater than 30 (see Section \ref{sec:New RB Classifier}) are considered plausible candidates and presented for visual inspection. If the number of plausible sources in a single CCD image exceeds 50, we reject all detections in that CCD because the quality of subtracted image is likely to be poor.

\subsection{Further follow-up mechanisms}
We manage data sets of transient candidates, building on a web user interface developed by \citet{Scalzo2017PASA...34...30S} based on the Django framework. Following the initial automated typing of the candidates, the next step requires a human to visually inspect and classify the remaining candidates in a web interface. All plausible candidates are assigned a unique name. Since we visit any field at least twice during the search, we prioritise candidates in the visual examination that are detected at least twice. We keep a list of "active" candidates, which human vetting has identified as relevant for spectroscopic or photometric follow-up. If another facility reports a kilonova discovery, we can add such an object manually. 

The web interface displays light curves, thumbnails of new, reference and subtracted images, and information from external services. Because of the rapid nature of the kilonova evolution and the robotic nature of SkyMapper operation, we added features to the web interface that trigger interaction with the telescope scheduler for follow-up modes that we anticipate to use during the next GW observing run (O4):
\begin{itemize}
    \item {\it Candidate Dweller:} this feature can be used to monitor one or several objects in one of three possible modes: (1) The "one-off" mode triggers just three exposures in the filter sequence \(i\)-\(u\)-\(i\) using an exposure sequence of 100-300-100 seconds (default but changeable). This mode is designed to probe the colour of a source quickly after an initial detection is made to help assess its likelihood of being a kilonova from a BNS merger. Two further modes with different sampling patterns can be used for continuous monitoring. (2) The "sampling" mode probes the lightcurve and its colour evolution with alternating 100\,sec \(i\) band and 300\,sec \(u\) band exposures. And (3) the "intense sampling" mode takes a series of ten consecutive 100\,sec \(i\)-band images followed by one 300\,sec \(u\)-band image, which provides the highest possible cadence, while recording colour evolution at lower cadence. Both sampling modes will continue observing the chosen target(s) until {\refbf it sets or the dome closes}. 
    
    \item {\it Alert Schedule Cleaner:} this feature simply removes all observations in the queue, aborting any previously committed sequence and allowing a fresh start.
    
    \item {\it AlertSDP Status:} this web page reports the execution status of individual pipeline tasks (Section \ref{subsec:AlertSDP}) and includes execution times for all jobs in the workflow.
\end{itemize} 
We also trigger prompt Target-of-Opportunity spectroscopy at the ANU 2.3m Telescope to verify the physical nature of relevant candidates and inform further SkyMapper activities.

\section{Ensemble-based transient classifier}
\label{sec:New RB Classifier}
In this section we describe a new ensemble-based approach for transient classification. The previous RF classifier had a completeness of $\sim 83$\% at 95\% purity, declining to $\sim 70$\% at 99\% purity (see \citealt{Scalzo2017PASA...34...30S}). The classifier was in need of retraining, because (i) the seeing range of images changed from the previous SkyMapper Transient Survey that mostly used bad-seeing time, and (ii) the characteristics of the image noise changed after implementing the SDP-style image calibration in the AlertSDP pipeline. However, the image sample used to train the RF classifier was not reprocessed with the AlertSDP and hence not available for retraining. Instead a new image set and transient sample was required. We used the opportunity given by the need to start from scratch to switch to a gradient boosted tree model (Section \ref{subsec:XGBoost}), as implemented in XGBoost \citep{Chen10.1145/2939672.2939785}, and develop a new training set from SMSS imagery (Section \ref{subsec:validation sample}). While comparing results with the performance of the previous RF classifier, we found that the two classifiers have complementary strengths and weaknesses. By combining the outputs of both classifiers, we were able to improve both purity (Section \ref{subsec:purity test}) and completeness (Section \ref{subsec:completeness test}). 

\subsection{New XGBoost Classifier}
\label{subsec:XGBoost}
We adopt XGBoost\footnote{\url{https://github.com/dmlc/xgboost}} using an ensemble of decision trees. It uses a gradient boosting algorithm \citep{Friedman00greedyfunction} to minimise a loss function when adding new decision trees. Unlike random forest methods, which train each tree independently, gradient boosted trees are built sequentially such that each subsequent tree aims to reduce the errors from its predecessors. The accuracy of classification is improved as more trees are added to the model, although a large complexity of the trees can lead to overfitting. However, XGBoost provides additional regularisation hyperparameters that can help reduce model complexity and guard against overfitting. Therefore, we use a column subsampling option to ensure that it uses a random subsample of the training data prior to growing trees. 

The robustness of the classifier model is mainly limited by the available training data of candidates with known class label (e.g., \citealt{Brink2013MNRAS.435.1047B, Scalzo2017PASA...34...30S}). The labelled data from the earlier supernova survey was unfortunately not useful in this regard, because it was predominantly obtained in bad seeing conditions. A new training set was generated by randomly selecting 1\,000 SMSS DR3 images in \(i\)-band, excluding low galactic latitudes of $\mid b\mid <15^{\circ}$. These data represent a large range of observational conditions and image quality from the survey. In this data set we searched for transient candidates and used the original RF classifier to filter the list of millions of candidates. We then eyeballed candidates with RBscore > 40, and labelled them as real or bogus, where the latter category includes bad subtractions, cosmic ray hits and warm pixels. We also added to the real set known asteroids, variable stars, quasars and a small number of candidates projected onto galaxies from the 6dFGS and 2MASS XSC catalogues, as these may appear like typical host galaxies of kilonovae. Finally, we added to the bogus set a random subset of candidates with RBscore < 30, provided they were not associated with known objects {\refbf of those types mentioned above that may be genuinely variable; this sample mostly includes subtraction artefacts such as residual features around bright stars, bad pixels and cosmic rays.} The real-labelled class has much fewer instances than the bogus-labelled class, with a number ratio close to 1:10. 

\begin{table}
\caption{Explored and chosen (bold) XGBoost Hyperparameters. Note that the results did not vary strongly with parameter changes.}
\centering
\begin{tabular}{lc}
\hline \noalign{\smallskip}
\noalign{\smallskip}
Hyperparameters & Values \\
\noalign{\smallskip} \hline \noalign{\smallskip}
\verb learning_rate &  0.1, 0.3, \textbf{0.5}, 0.7, 0.9\\ 
\verb max_depth & 3, \textbf{6}, 12, 18  \\ 
\verb n_estimators & 10, 50, 150, \textbf{300}  \\ 
\verb colsample_bytree &  0.1, 0.3, \textbf{0.5}, 0.7, 0.9\\ 
\verb lambda & \textbf{1}, 1.5, 3\\ 
\verb alpha &  \textbf{5}, 10, 15, 30\\ 
\noalign{\smallskip} \hline
\end{tabular}
\label{tab:tab1}
\end{table}

For training the XGBoost classifier with the similar input features used in \citet{Scalzo2017PASA...34...30S}, we split the sample into training and testing sets of 43\,593 and 71\,307 candidates, respectively. One issue is that our binary classification does not have a balanced number of instances in the training set. This requires the use of a stratified sampling strategy to learn the features of each class equally. Next, the classifier has a list of hyperparameters that require fine-tuning in order to derive the best-possible model. We select test values of hyperparameters from a set of points in a coarse grid (see Table \ref{tab:tab1}). We focused on hyperparameters that tend to have a high impact on the classification, such as control overfitting, learning rate, and complexity of the trees. Briefly the parameters are:

\begin{itemize}
    \item \verb learning_rate : step size shrinkage used in update to prevent over-fitting,
    \item \verb max_depth : maximum depth of a tree,
    \item \verb n_estimators : the number of trees in our ensemble,
    \item \verb colsample_bytree : the subsample ratio of columns when constructing each tree, 
    \item \verb lambda : L2 regularisation term on weights,
    \item \verb alpha: L1 regularisation term on weights.
\end{itemize} 
The final parameters were chosen as the model with the lowest classification error, i.e., with both high completeness and high purity (see bold figures in Table \ref{tab:tab1}).

\subsection{Validation sample} 
\label{subsec:validation sample}
To enable an accurate and unbiased assessment of classifier purity (Section \ref{subsec:purity test}), we need to define a "validation sample" -- a random sample of objects with the weight factors based on the observed magnitude and RB scores. In the general case of an unweighted random sample, high RB scores are less common than low ones. Also, it is required to keep reasonable statistics for the rare, bright objects. 
We note that the distribution of classes in the validation set is unbalanced and neither reflects those in the training set perfectly nor those expected during the real transient search. The role of the validation set is to lead the parameters of the classifier towards best performance on real data. Hence, the distribution of classes in the validation set should ideally reflect that of the classes in the test set, so that the performance metrics will be similar on both sets. In other words, the validation set should reflect the expected data imbalance. The imbalance in our validation set thus leads to suboptimal performance. In Section~\ref{sec:application of new metric} we test its performance in a real transient search during the O3 run.

From a set of 149 DR3 \(i\)-band images, we obtained 11,790 sources with RBscore larger than 30 without any further filtering, and we selected 4\,500 candidates for further visual inspection in a manner that sampled the range of source parameters. We label the sources irrespective of whether they have an automatic label or not. Table \ref{tab:tab2} summarises the steps we used to select pure transient candidates by removing known image artefacts (cleaned sample 1) and pre-selected categories (cleaned sample 2). Sample 1 can evaluate basic system performance, while sample 2 represents transient candidates that remain unexplained after automatic association with a known variable source and need to be presented for human vetting when hunting for real kilonovae. 

\begin{table}
\caption{Sample Selection Criteria for Purity Test}
\centering
\begin{tabular}{ccc}
\hline \noalign{\smallskip}
\noalign{\smallskip}
Sample & Selection Criteria & N\\
\noalign{\smallskip} \hline \noalign{\smallskip}
Raw & No filtering  & 4\,500 \\ 
\noalign{\smallskip} \hline \noalign{\smallskip}
& Known detector damages & 4\,490 \\ 
Cleaned & Residual CR hits + flagged pixels & 3\,519 \\ 
sample 1 & Subtraction artefacts & 2\,255 \\ 
& BrightStar-labelled sources & 2\,176\\
\noalign{\smallskip} \hline \noalign{\smallskip}
& Var-labelled sources & 2\,102 \\ 
Cleaned & Quasar-labelled sources & 2\,100 \\ 
sample 2 & Star-labelled sources & 1\,745 \\ 
& Asteroid-labelled sources & 332 \\  
\noalign{\smallskip} \hline
\end{tabular}
\label{tab:tab2}
\end{table}

Our main interest is in classifying either isolated (with no apparent host galaxy) or supernova-like transients with high completeness. To measure the classifier completeness (see Section \ref{subsec:completeness test}), we initially used a total of 5\,194 asteroid detections that were identified by SkyBot and 443 detections for 26 supernovae (16 SN\,Ia, 7 SN\,II, 1 SN\,IIn, 1 SN\,Ibc, and 1 SN\,Ic) that had been followed-up or discovered by the SkyMapper Transient Survey \citep{Scalzo2017PASA...34...30S,2019IAUS..339....3M}.

\begin{figure}[!t]
\centering
\includegraphics[width=\linewidth]{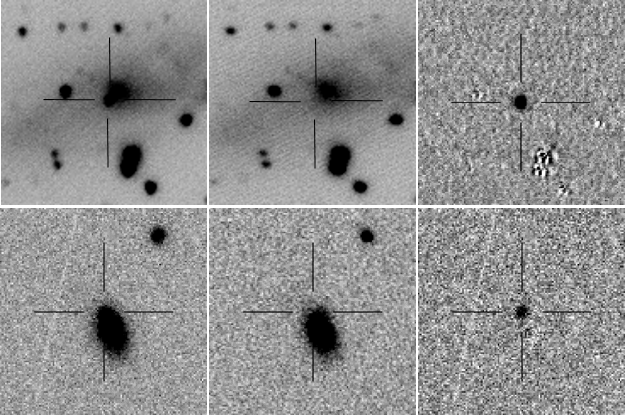}
\caption{Thumbnail images of two supernovae in the SMSS DR3 \(i\)-band dataset, showing the new, reference, and subtraction image from left to right {\refbf (size: 1 arcmin $\times$ 1 arcmin). Top: Type II SN\,2019ejj at d=13 Mpc, $i$ = 17.1. Bottom: Type Ia-91T SN\,2019ur at d=250 Mpc, $i$ = 18.8.}}
\label{figure: SkyMapper SNe in DR3}
\end{figure}

We additionally use the Open Supernova Catalogue \citep{Guillochon2017ApJ...835...64G} to collect spectroscopically confirmed SNe with distances less than 250 Mpc. In order to match known SNe against DR3, we take a broad date range of $\pm$30 days from the SN discovery date as our reference epoch. With this selection cut, there are 429 \(i\)-band images of 318 SNe, where the position and epoch were matched with sources in the SMSS DR3 catalogue within 5 arcseconds. Some of those DR3 sources are the host galaxies of SNe, or otherwise chance superpositions of unrelated sources. After running the image set through the AlertSDP pipeline, we recover 153 detections for 117 SNe which lie in the magnitude range between 15.5 and 20.5 in \(i\) band. Since the SMSS is focused on covering the entirety of the southern sky rather than frequently repeating a given area, most of the SNe only appear in a single exposure. By type, this sample contains 69 Type I SN (64 Ia, 2 Ib, 2 Ic, 1 Ib/Ic), 45 Type II SN (36 II, 4 IIP, 2 IIb, 3 IIn), and 3 unclassified ones, in a variety of host galaxies. Thus, our completeness test with the Open Supernova Catalog sample will be less affected by host galaxy selection effects than that of the SMT SNe sample. Figure \ref{figure: SkyMapper SNe in DR3} shows thumbnail images of representative SN examples in different redshift ranges. We assume that this validation sample contains a more representative sample of contaminants, but also represents the outcome of the pipeline more accurately than the training data used in the model development stage.

\begin{figure*}[!t]
\centering
\includegraphics[width=0.497\textwidth]{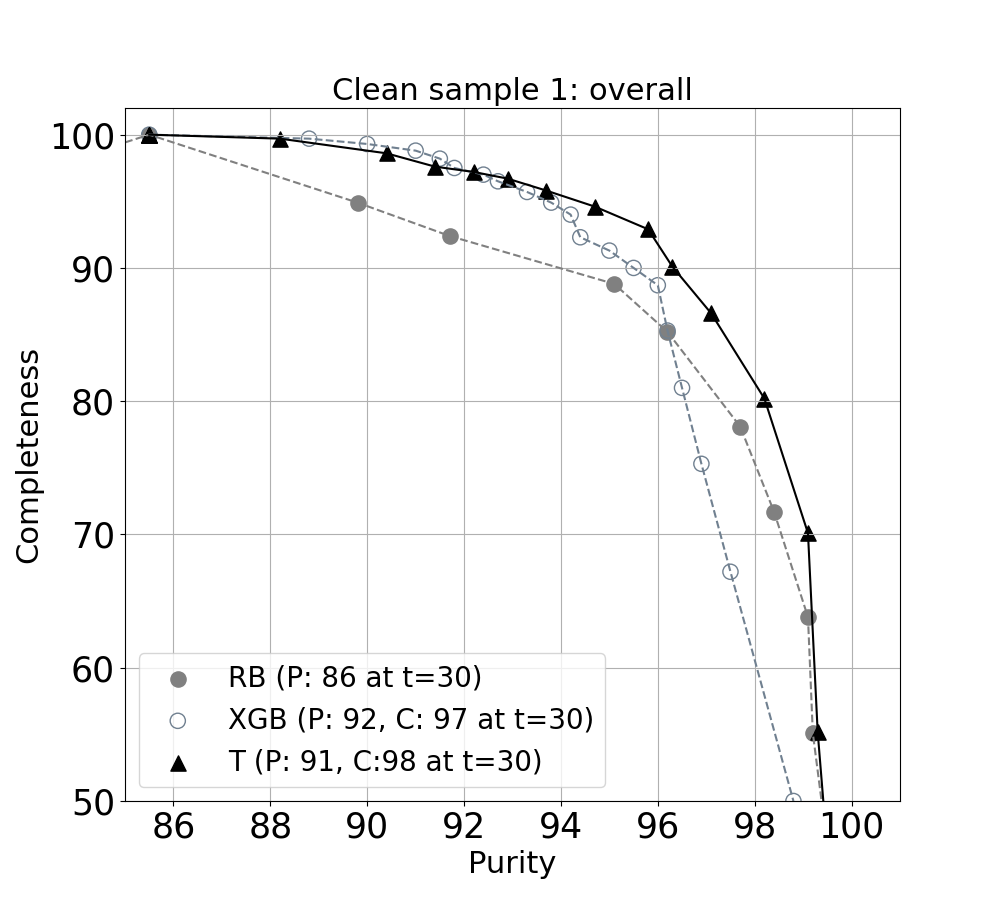}
\includegraphics[width=0.497\textwidth]{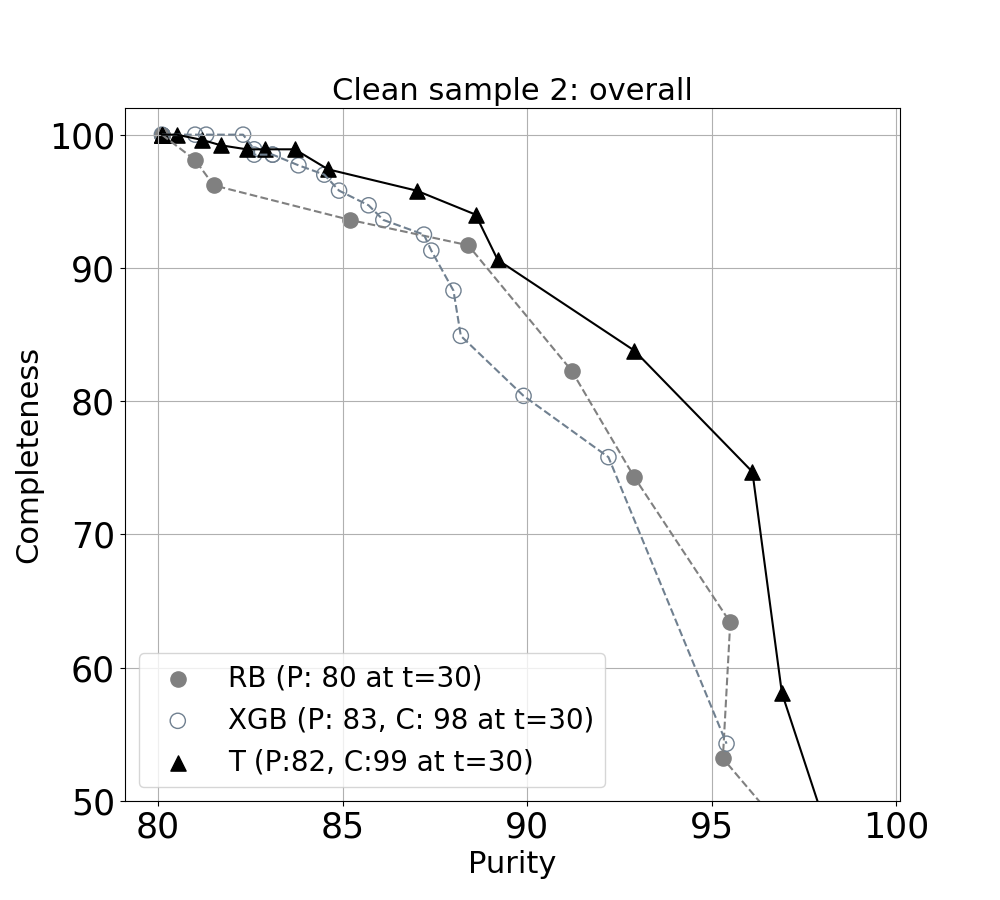}
\caption{Purity vs. completeness curves for the two cleaned samples described in Section \ref{subsec:validation sample}. We compare the performance of three ensemble scores: RBscore (grey filled circle), XGBscore (grey open circle), and our new metric, Tscore (black filled triangle). The text in the panels refers to purity (P) and completeness (C) scores at a threshold of $t=30$, which we adopt for our transient search.}
\label{figure: FANT4STIC}
\end{figure*}

\subsection{Purity test}
\label{subsec:purity test}
As our two classifiers have different strengths, a combination of them can lead to better performance. XGBoost is much better at recognising bad subtractions with negative pixel values, whereas RF works better at recognising that CR hits and warm pixels are not real objects. While this may be the result of different training, we define a new metric here, Tscore, that combines the two machine learning models (known as an ensemble of classifiers; see Chapter 34 in \citealt{Dietterich2000} for instance), using the simple rule:

$$ {\rm{Tscore}} =  \frac{\rm RBscore + \rm XGBscore}{2}  ~,$$ 
which gives both classifiers a similar weight.

For any threshold $t$, we define classifier purity:
$$ \mathrm{Purity}= \frac{\rm TP}{\rm TP + \rm FP} ~,
$$ where TP is the number of true transients with any scores greater than $t$, and FP is the number of false positives that were incorrectly considered as real. Varying the threshold of a binary classifier usually trades off better completeness against better purity. In Figure \ref{figure: FANT4STIC}, we show curves of completeness vs. purity in our cleaned samples as a function of threshold and compare our two classifiers as well as the combined Tscore. In cleaned sample 1, XGBscore is an improvement over RBscore at every completeness $>90$\%; for further applications, we choose a threshold of $t=30$, which delivers 92\% purity and 97\% completeness. However, at higher purity the originally used RBscore appears more complete. Our new metric then combines the advantages of both and is in all parts of the curve at least as good as either the RB or XGB classifier. In cleaned sample 2 (right panel of Figure \ref{figure: FANT4STIC}) all transient candidates explained by known objects have been removed and only those in need of human inspection are left over; here, a threshold of $t=30$ is still a good choice and provides a purity above 80\% with 99\% completeness. Based on this, we choose the ensemble-based Tscore classifier with a threshold of $t=30$ as our final classifier. {\refbf Other wide-field surveys have obtained similar results on real-bogus classification by using convolutional neural networks 
\citep[e.g.][]{Duev2019MNRAS.489.3582D, 2021arXiv210209892K}.}

\begin{figure*}[!t]
\centering
\includegraphics[width=0.33\linewidth]{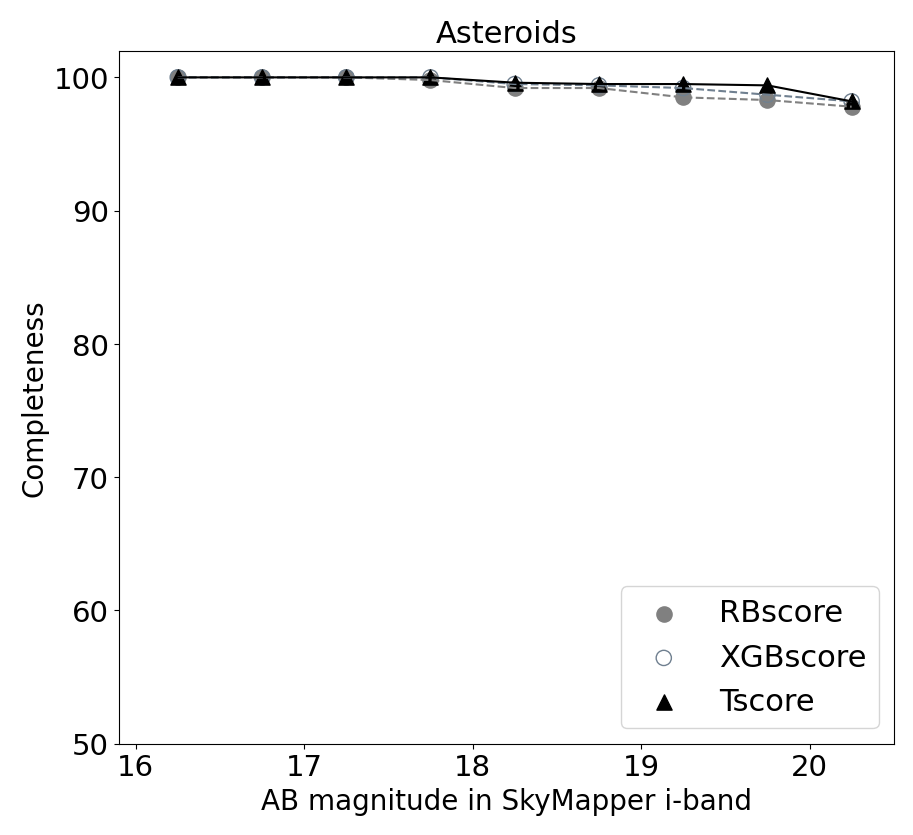}
\includegraphics[width=0.33\linewidth]{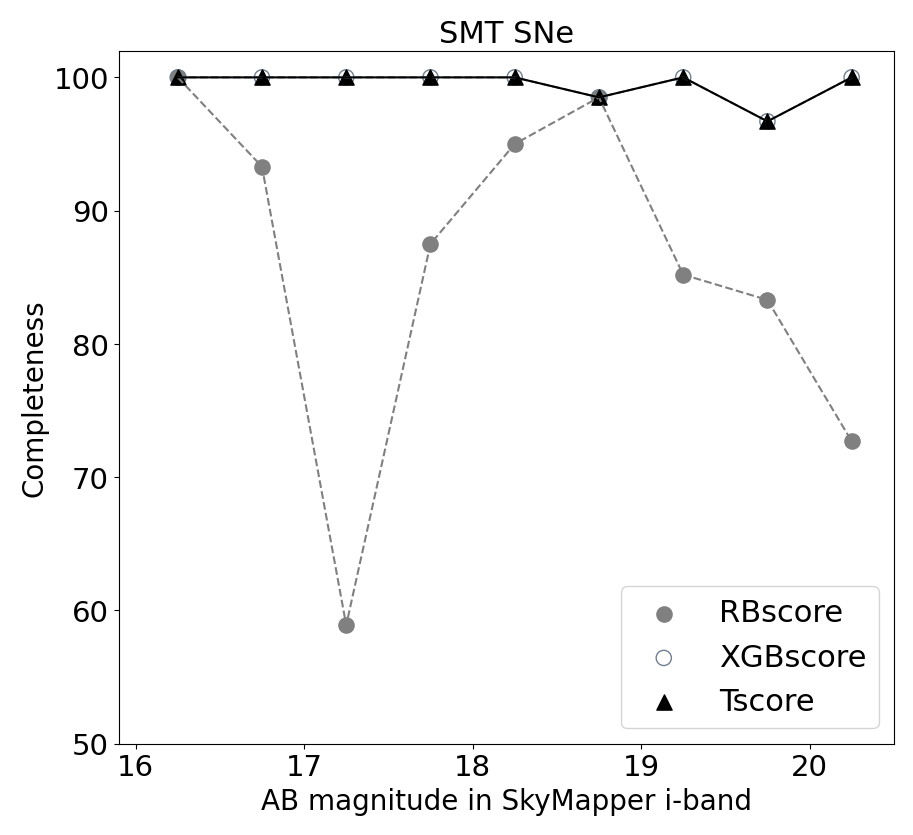}
\includegraphics[width=0.33\linewidth]{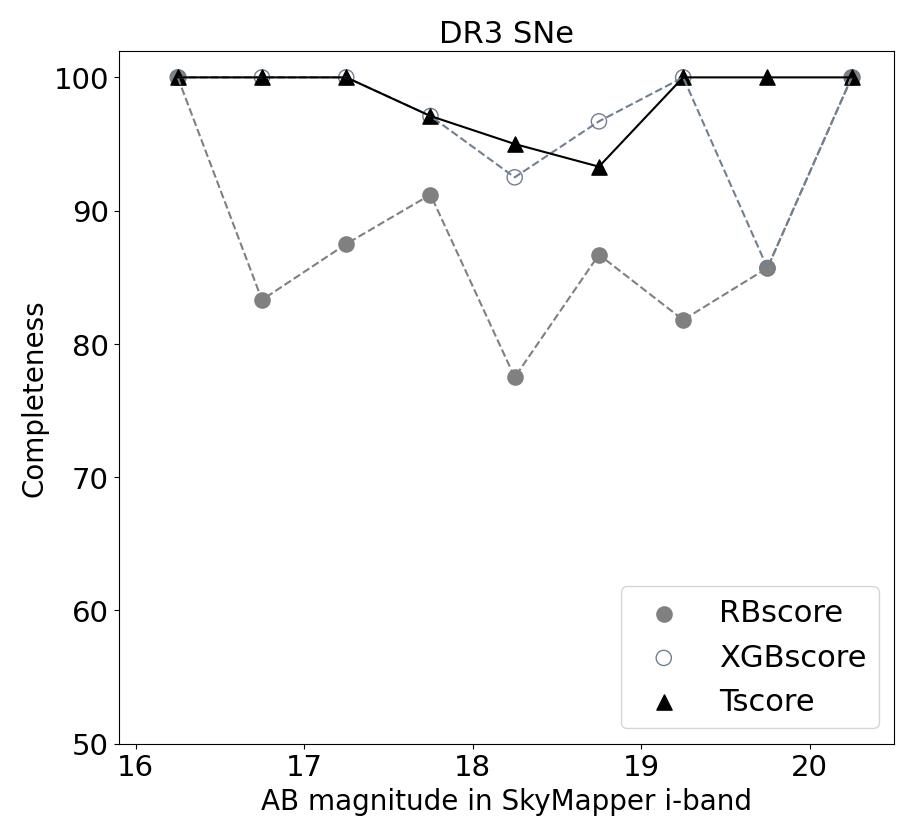}
\caption{Completeness test with asteroid (left), SMT SN (middle), and DR3 SN (right) samples as a function of magnitude. We compare the three different ensemble scores: RBscore (grey filled circle), XGBscore (grey open circle), and Tscore (black filled triangle). The same axis ranges are used in each panel.}
\label{figure: Completeness test}
\end{figure*}

\subsection{Completeness test}
\label{subsec:completeness test} 
Next, we check the completeness of Tscore classifier for real-bogus classification. For any threshold $t$ on the score, we obtain the completeness or recovery rate as a function of magnitude, using 
$$
\mathrm{Completeness} = \frac{\rm TP}{\rm TP + FN} ~,
$$ where TP is the number of true positives in the test set that were correctly classified as real and FN is the number of false negatives (= positives misclassified as bogus). Figure \ref{figure: Completeness test} shows how the classifier performs as a function of magnitude for the two cases of asteroids and supernovae. In both cases, there is a clear trend for brighter transients to have a higher recovery rate at a given threshold. The high recovery rate of asteroids can be explained by the fact that they are over-represented in our training set compared to SN-like transients. Asteroids also appear mostly as isolated sources and are only rarely blended with galaxies, while supernovae and kilonovae are mostly blended with their host galaxy. With a threshold of $t=30$, we attain 99.5\% completeness for both asteroid (TP=5\,170, FN=24) and supernova (TP=441, FN=2) classifications. While the SN sample has 443 detection images, these are from only 26 host galaxies and hence its statistical significance is smaller than it may seem. Moreover, there is reason to expect variations in completeness as a function of SN magnitude (at the time of detection), SN offset from the host galaxy nucleus, morphological type of the host galaxy, and redshift.

To overcome this weakness, we test the completeness also for the DR3 SNe sample, which has fewer images but more (117 vs. 26) host galaxies. Using $t=30$ again, we find high completeness ($\sim$97\%; TP=148, FN=5) to the lowest magnitudes, {\refbf comparable to other recently reported results \citep[with $\sim $97\%]{2021arXiv210209892K}.} We checked the fives failures of the classifier and noticed that in three cases the FNs are superimposed on the host nucleus within $<$ 1 arcsecond offset, so that the classification may be affected by residual features from poor subtraction. The two SNe outside the nuclei of their hosts are OGLE\,16etd, a classical SN II found close to maximum light in a crowded region, and SN\,2018aau, an apparently hostless SN Ia detected $\sim$2 weeks after maximum. The latter has two observations, one classified as TP and another one as FN. The classification scores of a given transient candidate can indeed vary with seeing conditions and star density in the field, which affect image subtraction.

\begin{table*}[!t]
\small
\caption{Summary of preliminary BNS detection alerts in O3. For GW190425, we list updated information from \citet{2020ApJ...892L...3A}.}
\centering
\begin{tabular}{lccccccccl}
\hline \noalign{\smallskip}  
&  \multicolumn{2}{c}{Preliminary Area} & &  &  \multicolumn{3}{c}{$P_\mathrm{class}$ preliminary (updated)} & &\\
\noalign{\smallskip} \cline{2-3} \cline{6-8} \noalign{\smallskip} 
Superevent & 50\% & 90\% & D$_{L}$ & FAR & BNS & NSBH & Terrestrial & This & Notes\\
 (ID)  & (deg$^2$) & (deg$^2$) & (Mpc) & (yr$^{-1}$) & (\%)  & (\%) & (\%) & work &    \\
\noalign{\smallskip} \hline \noalign{\smallskip}
GW190425 &     & 8284 & 159$^{+69}_{-71}$ & 1 per 69\,000 & 100 & - & - & \checkmark & \citet{2020ApJ...892L...3A} \\
S190426c  & 472 & 1932 & 377$\pm$100 & 1 per 1.63 & 49 (24) & 13 (6) & 14 (58) &  & Too distant to detect\\ 
S190510g  & 575  & 3462 & 277$\pm$92 & 1 per 3.59  & 98 (42) & 0 (0) & 2 (58) & \checkmark & Cloudy weather\\ 
S190901ap & 4176 & 13613  & 241$\pm$79 & 1 per 4.51 & 86 & 14 & 0 &  & Poor localisation\\
S190910h  & 8066 & 24226 & 230$\pm$88 & 1 per 0.88 & 61 & 0 & 39 &   & Poor localisation\\
S191213g  & 259 & 1393 &201$\pm$81& 1 per 0.89 & 77 & 0 & 23 &  & Near the Sun\\
S191220af & 580 & 5238  & 125$\pm$28 & 1 per 79.96 & $>$99 & 0 & $<$1& \checkmark & Retracted \\
S200213t  & 150 & 2587 & 210$\pm$80& 1 per 1.79 & 63 & 0  & 37 & & Marginal FAR \\
\noalign{\smallskip} \hline
\noalign{\smallskip}
GW170817 &  & 28 & 40$^{+8}_{-14}$ & 1 per 80\,000 & 100 & - & - & & \citet{PhysRevLett.119.161101}\\
\noalign{\smallskip}
\noalign{\smallskip} \hline
\end{tabular}
\label{tab:tab3}
\end{table*}

\section{Application in the LIGO/Virgo O3 run}
\label{sec:application of new metric}
Our follow-up program was dedicated to the rapid search for kilonova counterparts to BNS mergers beyond GW170817 that would be observed by LIGO/Virgo observatories in O3. To trigger the SkyMapper observations, the estimated mass of one component of the compact binary system is required to be consistent with a neutron star (via the HasNS probability in the GCN notice). Table \ref{tab:tab3} summarises BNS candidates that have been reported in the LIGO/Virgo O3 public alerts, including events retracted later. For comparison, we list the properties of GW170817 detected in the O2 campaign in the last row of Table \ref{tab:tab3}. 

SkyMapper responded to three of these triggers, S190425z, S190510g and S191220af. The other triggers were ignored because of poor localisation or proximity to the Sun. The most promising event in all of O3 was S190425z, also called GW190425, where SkyMapper managed to obtain a data set of typical sky coverage. While we responded to the event in real-time (see Section \ref{subsec:real-world event}), we also reprocessed the data with the latest version of the AlertSDP to evaluate the real-world performance of the pipeline (see Section \ref{subsec:real-world performance}). 

\subsection{The search for an EM counterpart to S190425z/GW190425} 
\label{subsec:real-world event}
On 25 April 2019 08:18:05.017 (UT), the LIGO Livingston observatory alone identified a GW chirp signal with a false alarm rate of one per 69\,000 years and a signal-to-noise ratio (SNR) of 12.9 \citep{2020ApJ...892L...3A}. The LIGO Hanford facility was offline when the event was detected and the Virgo facility did not contribute to its detection due to a low SNR=2.5. This signal has strong evidence for the mass of one or both components to be consistent with a neutron star (HasNS$=$100\%; BNS$=$100\%). It also shows a clear time-frequency map with the characteristic upwardly-sweeping chirp pattern expected for an inspiralling BNS system, which is what had been seen in GW170817 (see Figure 1 of \citealt{PhysRevLett.119.161101}). The latest inferred position of this event was poorly localised in the sky, with a 90\% credible sky area covering 8\,284 deg$^{2}$, but the distance inferred from the GW signal was $159^{+69}_{-71}$ Mpc (see \citealt{2020ApJ...892L...3A} for details). Unlike the case of GW170817, where the position constraints were $\sim 250\times$ tighter, this situation makes {\refbf it difficult to find faint EM counterparts quickly. While the extremely wide-field Zwicky Transient Facility (ZTF) did search $\sim 8 000$ deg$^2$ and reported two potential candidates, they needed two nights of observing to cover the area \citep{2019ApJ...885L..19C}. Some searches even reached sufficient depth to detect GW170817-like kilonovae, although their areal coverage was incomplete \citep{2019ApJ...880L...4H}.}



The first optical detection by SkyMapper was made from the prioritised list of target fields that were observed about 6 hours after the merger. We let the search mode (Section \ref{subsec:alert response}) run on a relatively small fraction of the southern localisation region ($\sim$125 square degrees), acknowledging a small probability of finding the associated EM counterpart, as the area covered only $\sim$1\% of the initial BAYESTAR map and $\sim$3\% of its integrated probability. We found two transient candidates that had two detections separated by about 10 minutes, but neither of them had visible host galaxies. After the discovery of two candidates by the Zwicky Transient Facility (ZTF) survey, we moved onto a new phase to help determine whether either transient might be related to the GW signal. In Section \ref{subsec:real-world performance}, we perform a simple experiment to test our new software implementation with the SkyMapper observations of the BNS merger GW190425. 

As part of our Target-of-Opportunity programme at the ANU 2.3m Telescope, we obtained a series of early spectra of the potential counterpart ZTF19aarykkb {\refbf  \citep{2019ApJ...885L..19C}}. We used the WiFeS instrument \citep{2007Ap&SS.310..255D}, whose integral-field nature allows us to obtain spectra of both the transient and the host galaxy simultaneously. These could identify a kilonova from its spectral features being atypical for better-known transient types (e.g., cataclysmic variable stars or supernovae) and would also provide constraints on the environment of the transient as a birth or merger place for the progenitor binary (e.g., \citealt{Levan2017ApJ...848L..28L}). We obtained our first 850\,s exposure starting at 2019-04-25 17:40:40 UT, followed by 5$\times$1200s spectra, using the $R\sim3000$ gratings, B3000+R3000, covering the wavelength range of 320nm to 990nm. The WiFeS spectra showed the bright ZTF candidate (\(r\)= 18.63\,mag) having spectral features similar to those of a young Type II supernova \citep{Chang2019GCN.24260....1C}.

A number of additional EM candidates were discovered by other facilities {\refbf  (e.g., \citealt{2019ApJ...880L...4H})}. At the location (RA=17:02:19.2, Dec=-12:29:08.2) of the Swift UVOT candidate by \citet{Breeveld2019GCN.24296....1B} we did not see any transient down to an $i=20$ at the two epochs: 2019-04-25 14:17:00 and 2019-04-25 14:25:53 UT \citep{Chang2019GCN.24325....1C}. Later work suggested the Swift detection was likley to be an ultraviolet flare on an M dwarf star \citep{Lipunov2019GCN.24326....1L,Bloom2019GCN.24337....1B}. As discussed by previous studies, M dwarf flares are the most common transient phenomenon that may confuse searches for kilonova candidates from GW events (e.g., \citealt{vanRoestel2019MNRAS.484.4507V, Chang2020MNRAS.491...39C}).

\subsection{SkyMapper search results for GW190425} 
\label{subsec:real-world performance}
To get a realistic view of the current performance of the SkyMapper transient facility, we reran the initial search mode data from the first night of the GW190425 event in O3 through the AlertSDP as described above. Each target field in the $\sim$125 square degrees of the localisation region was visited twice, separated by about 10 minutes.
We applied the same filtering scheme for the selection of transient candidates as introduced in Section \ref{sec:New RB Classifier}. Table \ref{tab:tab5} shows how each filtering step reduced the number of candidates from 41\,497 to 359, which is a >100-fold reduction in the number of candidates requiring human inspection. Before the vetting process, this number was reduced even further to 9 candidates (1 in $\sim$5,000) by considering only those with two detections. Each of the remaining candidates could be one of the following categories: (i) an isolated transient in a position which was not reported before, (ii) a galaxy with a SN at the time of observation, (iii) an as-yet unidentified variable object that has historical detections, or (iv) a bogus detection. We note that there was one known SN that was not detected in our search, because it was fainter than our detection limit ($i$ = 20.4). Five of the sources were clear bogus detections caused by poor image subtraction. The remaining four candidates are persistent point sources in SMSS DR3 and seen to vary by 0.2 to 0.7 mag over the 5-year span of data in DR3. Thus, we conclude that none of {\refbf them were related to GW190425}.

\begin{table}
\caption{Transient selection for GW190425}
\centering
\begin{tabular}{cc}
\hline \noalign{\smallskip}
\noalign{\smallskip}
Selection Criteria & N\\
\noalign{\smallskip} \hline \noalign{\smallskip}
No filtering & 41\,497\\
Tscore$>$30 & 6\,619 \\ 
\noalign{\smallskip} \hline \noalign{\smallskip}
All flagged pixels & 4\,345\\
Subtraction artefacts & 4\,131 \\ 
BrightStar-labelled sources & 3\,067\\
\noalign{\smallskip} \hline \noalign{\smallskip}
Var-labelled sources & 2\,694 \\ 
Quasar-labelled sources & 2\,689 \\ 
Star-labelled sources & 825 \\ 
Asteroid-labelled sources & 359 \\ 
\noalign{\smallskip} \hline \noalign{\smallskip}
Candidates detected twice & 9\\ 
\noalign{\smallskip} \hline
\end{tabular}
\label{tab:tab5}
\end{table}

\section{Summary and outlook to O4}
\label{sec:summary}
In this paper, we describe the capability of the SkyMapper facility for rapid-response EM observations of GW events. Developed as a new branch of the SkyMapper pipeline, the AlertSDP provides a prompt response to GW alerts with low-latency data processing and a web interface for follow-up. With a hybrid combination of ensemble-based classifiers, we achieve a high completeness ($\sim$98\%) and purity ($\sim$91\%) in real-bogus classification; this is measured with a sample of 4\,500 random transient candidates in the purity test, and over 5\,000 asteroids as well as 143 supernovae in the completeness test. We performed the same analysis on the real SkyMapper data for LIGO/Virgo event S190425z/GW190425, confirming its efficiency to reduce contaminants. 

When the observing run O4 starts with the GW detector network of the LIGO/Virgo/Kagra (LVK) collaboration, the predicted rate of BNS merger detections is 10$^{+52}_{-10}$ per year, their predicted median luminosity distance is 170$^{+6.3}_{-4.8}$ Mpc, and their predicted median 90\% sky localisation area is 33$^{+4.9}_{-5.3}$ deg$^{2}$ {\refbf\citep[all values given as 5\% to 95\% confidence intervals; ][]{ 2020LRR....23....3A}}. A 100-sec $i$-band exposure with SkyMapper can discover {\refbf an GW170817-like kilonova} at four times the distance of GW170817. Future SMSS data releases will have deeper co-added reference images with up to 600\,sec exposures per sky pixel in each band, which may allow complete detection of transient candidates to $i\approx 21$ mag and provide good coverage beyond the LVK median horizon of 170 Mpc. The AlertSDP should identify all transients in a $\sim 100$ deg$^{2}$ search area within four hours from the start of observations. {\refbf In the future}, we expect to reduce this latency by speeding up the data processing. 

\section*{Acknowledgements}
{\refbf We are grateful to an anonymous referee for comments that significantly improved the manuscript.} We thank Jon Nielsen (ANU) for parallelising part of the AlertSDP. This research was partly funded by the Australian Research Council Centre of Excellence for Gravitational Wave Discovery (OzGrav), CE170100004. SWC acknowledges support from the National Research Foundation of Korea (NRF) grant, No. 2020R1A2C3011091, funded by the Korea government (MSIT). CAO acknowledges support from the Australian Research Council through Discovery Project DP190100252. The national facility capability for SkyMapper has been funded through ARC LIEF grant LE130100104 from the Australian Research Council, awarded to the University of Sydney, the Australian National University, Swinburne University of Technology, the University of Queensland, the University of Western Australia, the University of Melbourne, Curtin University of Technology, Monash University and the Australian Astronomical Observatory. SkyMapper is owned and operated by The Australian National University's Research School of Astronomy and Astrophysics. The survey data were processed and provided by the SkyMapper Team at ANU. The SkyMapper node of the All-Sky Virtual Observatory (ASVO) is hosted at the National Computational Infrastructure (NCI). Development and support of the SkyMapper node of the ASVO has been funded in part by Astronomy Australia Limited (AAL) and the Australian Government through the Commonwealth's Education Investment Fund (EIF) and National Collaborative Research Infrastructure Strategy (NCRIS), particularly the National eResearch Collaboration Tools and Resources (NeCTAR) and the Australian National Data Service Projects (ANDS). 

\bibliographystyle{pasa-mnras}
\bibliography{O3paper_Skymapper}

\begin{thebibliography}{}
\makeatletter
\relax
\def\mn@urlcharsother{\let\do\@makeother \do\$\do\&\do\#\do\^\do\_\do\%\do\~}
\definecolor{darkblue}{rgb}{0,0,0.597656}
\def\mndoi{\begingroup\mn@urlcharsother \@ifnextchar [ {\mndoi@} {\mndoi@[]}}
\def\mndoi@[#1]#2{\def\@tempa{#1}\ifx\@tempa\@empty \href
  {http://dx.doi.org/#2} {\textcolor{darkblue}{doi:#2}}\else \href
  {http://dx.doi.org/#2} {\textcolor{darkblue}{#1}}\fi \endgroup}
\def\mn@eprint#1#2{\mn@eprint@#1:#2::\@nil}
\def\mn@eprint@arXiv#1{\href {http://arxiv.org/abs/#1} {{\tt arXiv:#1}}}
\def\mn@eprint@dblp#1{\href {http://dblp.uni-trier.de/rec/bibtex/#1.xml}
  {dblp:#1}}
\def\mn@eprint@#1:#2:#3:#4\@nil{\def\@tempa {#1}\def\@tempb {#2}\def\@tempc
  {#3}\ifx \@tempc \@empty \let \@tempc \@tempb \let \@tempb \@tempa \fi \ifx
  \@tempb \@empty \def\@tempb {arXiv}\fi \@ifundefined
  {mn@eprint@\@tempb}{\@tempb:\@tempc}{\expandafter \expandafter \csname
  mn@eprint@\@tempb\endcsname \expandafter{\@tempc}}}

\bibitem[\protect\citeauthoryear{{Abbott} et~al.,}{{Abbott}
  et~al.}{2017a}]{PhysRevLett.119.161101}
{Abbott} B.~P.,  et~al., 2017a, \mndoi [\prl] {10.1103/PhysRevLett.119.161101},
  \href {https://ui.adsabs.harvard.edu/abs/2017PhRvL.119p1101A} {119, 161101}

\bibitem[\protect\citeauthoryear{{Abbott} et~al.,}{{Abbott}
  et~al.}{2017b}]{Abbott2017ApJ...848L..12A}
{Abbott} B.~P.,  et~al., 2017b, \mndoi [\apjl] {10.3847/2041-8213/aa91c9},
  \href {https://ui.adsabs.harvard.edu/abs/2017ApJ...848L..12A} {848, L12}

\bibitem[\protect\citeauthoryear{{Abbott} et~al.,}{{Abbott}
  et~al.}{2017c}]{2017ApJ...850L..40A}
{Abbott} B.~P.,  et~al., 2017c, \mndoi [\apjl] {10.3847/2041-8213/aa93fc},
  \href {https://ui.adsabs.harvard.edu/abs/2017ApJ...850L..40A} {850, L40}

\bibitem[\protect\citeauthoryear{{Abbott} et~al.,}{{Abbott}
  et~al.}{2020a}]{2020arXiv201014527A}
{Abbott} R.,  et~al., 2020a, arXiv e-prints, \href
  {https://ui.adsabs.harvard.edu/abs/2020arXiv201014527A} {p. arXiv:2010.14527}

\bibitem[\protect\citeauthoryear{{Abbott} et~al.,}{{Abbott}
  et~al.}{2020b}]{2020LRR....23....3A}
{Abbott} B.~P.,  et~al., 2020b, \mndoi [Living Reviews in Relativity]
  {10.1007/s41114-020-00026-9}, \href
  {https://ui.adsabs.harvard.edu/abs/2020LRR....23....3A} {23, 3}

\bibitem[\protect\citeauthoryear{{Abbott} et~al.,}{{Abbott}
  et~al.}{2020c}]{2020ApJ...892L...3A}
{Abbott} B.~P.,  et~al., 2020c, \mndoi [\apjl] {10.3847/2041-8213/ab75f5},
  \href {https://ui.adsabs.harvard.edu/abs/2020ApJ...892L...3A} {892, L3}

\bibitem[\protect\citeauthoryear{{Adhikari}, {Fishbach}, {Holz}, {Wechsler}  \&
  {Fang}}{{Adhikari} et~al.}{2020}]{2020ApJ...905...21A}
{Adhikari} S.,  {Fishbach} M.,  {Holz} D.~E.,  {Wechsler} R.~H.,   {Fang} Z.,
  2020, \mndoi [\apj] {10.3847/1538-4357/abbfb7}, \href
  {https://ui.adsabs.harvard.edu/abs/2020ApJ...905...21A} {905, 21}

\bibitem[\protect\citeauthoryear{{Alard} \& {Lupton}}{{Alard} \&
  {Lupton}}{1998}]{Alard1998ApJ...503..325A}
{Alard} C.,  {Lupton} R.~H.,  1998, \mndoi [\apj] {10.1086/305984}, \href
  {https://ui.adsabs.harvard.edu/abs/1998ApJ...503..325A} {503, 325}

\bibitem[\protect\citeauthoryear{{Andreoni} et~al.,}{{Andreoni}
  et~al.}{2017}]{Andreoni2017PASA...34...69A}
{Andreoni} I.,  et~al., 2017, \mndoi [\pasa] {10.1017/pasa.2017.65}, \href
  {https://ui.adsabs.harvard.edu/abs/2017PASA...34...69A} {34, e069}

\bibitem[\protect\citeauthoryear{{Arcavi}}{{Arcavi}}{2018}]{2018ApJ...855L..23A}
{Arcavi} I.,  2018, \mndoi [\apjl] {10.3847/2041-8213/aab267}, \href
  {https://ui.adsabs.harvard.edu/abs/2018ApJ...855L..23A} {855, L23}

\bibitem[\protect\citeauthoryear{{Arcavi} et~al.,}{{Arcavi}
  et~al.}{2017}]{Arcavi2017Natur.551...64A}
{Arcavi} I.,  et~al., 2017, \mndoi [\nat] {10.1038/nature24291}, \href
  {https://ui.adsabs.harvard.edu/abs/2017Natur.551...64A} {551, 64}

\bibitem[\protect\citeauthoryear{{Bailey}, {Aragon}, {Romano}, {Thomas},
  {Weaver}  \& {Wong}}{{Bailey} et~al.}{2007}]{Bailey2007ApJ...665.1246B}
{Bailey} S.,  {Aragon} C.,  {Romano} R.,  {Thomas} R.~C.,  {Weaver} B.~A.,
  {Wong} D.,  2007, \mndoi [\apj] {10.1086/519832}, \href
  {https://ui.adsabs.harvard.edu/abs/2007ApJ...665.1246B} {665, 1246}

\bibitem[\protect\citeauthoryear{{Barnes} \& {Kasen}}{{Barnes} \&
  {Kasen}}{2013}]{Barnes2013ApJ...775...18B}
{Barnes} J.,  {Kasen} D.,  2013, \mndoi [\apj] {10.1088/0004-637X/775/1/18},
  \href {https://ui.adsabs.harvard.edu/abs/2013ApJ...775...18B} {775, 18}

\bibitem[\protect\citeauthoryear{{Becker}}{{Becker}}{2015}]{Becker2015ascl.soft04004B}
{Becker} A.,  2015, {HOTPANTS: High Order Transform of PSF ANd Template
  Subtraction} (\mn@eprint {ascl} {1504.004})

\bibitem[\protect\citeauthoryear{{Berthier}, {Vachier}, {Thuillot}, {Fernique},
  {Ochsenbein}, {Genova}, {Lainey}  \& {Arlot}}{{Berthier}
  et~al.}{2006}]{Berthier2006ASPC..351..367B}
{Berthier} J.,  {Vachier} F.,  {Thuillot} W.,  {Fernique} P.,  {Ochsenbein} F.,
   {Genova} F.,  {Lainey} V.,   {Arlot} J.~E.,  2006, in {Gabriel} C.,
  {Arviset} C.,  {Ponz} D.,   {Enrique} S.,  eds,  Astronomical Society of the
  Pacific Conference Series Vol. 351, Astronomical Data Analysis Software and
  Systems XV. p.~367

\bibitem[\protect\citeauthoryear{{Bertin} \& {Arnouts}}{{Bertin} \&
  {Arnouts}}{1996}]{Bertin1996A&AS..117..393B}
{Bertin} E.,  {Arnouts} S.,  1996, \mndoi [\aaps] {10.1051/aas:1996164}, \href
  {https://ui.adsabs.harvard.edu/abs/1996A&AS..117..393B} {117, 393}

\bibitem[\protect\citeauthoryear{{Bertin}, {Mellier}, {Radovich}, {Missonnier},
  {Didelon}  \& {Morin}}{{Bertin} et~al.}{2002}]{2002ASPC..281..228B}
{Bertin} E.,  {Mellier} Y.,  {Radovich} M.,  {Missonnier} G.,  {Didelon} P.,
  {Morin} B.,  2002, in {Bohlender} D.~A.,  {Durand} D.,   {Handley} T.~H.,
  eds,  Astronomical Society of the Pacific Conference Series Vol. 281,
  Astronomical Data Analysis Software and Systems XI. p.~228

\bibitem[\protect\citeauthoryear{{Bessell}, {Bloxham}, {Schmidt}, {Keller},
  {Tisserand}  \& {Francis}}{{Bessell}
  et~al.}{2011}]{Bessell2011PASP..123..789B}
{Bessell} M.,  {Bloxham} G.,  {Schmidt} B.,  {Keller} S.,  {Tisserand} P.,
  {Francis} P.,  2011, \mndoi [\pasp] {10.1086/660849}, \href
  {https://ui.adsabs.harvard.edu/abs/2011PASP..123..789B} {123, 789}

\bibitem[\protect\citeauthoryear{{Bloom} et~al.,}{{Bloom}
  et~al.}{2012}]{Bloom2012PASP..124.1175B}
{Bloom} J.~S.,  et~al., 2012, \mndoi [\pasp] {10.1086/668468}, \href
  {https://ui.adsabs.harvard.edu/abs/2012PASP..124.1175B} {124, 1175}

\bibitem[\protect\citeauthoryear{{Bloom}, {Zucker}, {Schlafly}, {Finkbeiner},
  {Martinez-Palomera}, {Goldstein}  \& {Andreoni}}{{Bloom}
  et~al.}{2019}]{Bloom2019GCN.24337....1B}
{Bloom} J.~S.,  {Zucker} C.,  {Schlafly} E.,  {Finkbeiner} D.,
  {Martinez-Palomera} J.,  {Goldstein} D.~A.,   {Andreoni} I.,  2019, GRB
  Coordinates Network, \href
  {https://ui.adsabs.harvard.edu/abs/2019GCN.24337....1B} {24337, 1}

\bibitem[\protect\citeauthoryear{{Breeveld} et~al.,}{{Breeveld}
  et~al.}{2019}]{Breeveld2019GCN.24296....1B}
{Breeveld} A.~A.,  et~al., 2019, GRB Coordinates Network, \href
  {https://ui.adsabs.harvard.edu/abs/2019GCN.24296....1B} {24296, 1}

\bibitem[\protect\citeauthoryear{{Brink}, {Richards}, {Poznanski}, {Bloom},
  {Rice}, {Negahban}  \& {Wainwright}}{{Brink}
  et~al.}{2013}]{Brink2013MNRAS.435.1047B}
{Brink} H.,  {Richards} J.~W.,  {Poznanski} D.,  {Bloom} J.~S.,  {Rice} J.,
  {Negahban} S.,   {Wainwright} M.,  2013, \mndoi [\mnras]
  {10.1093/mnras/stt1306}, \href
  {https://ui.adsabs.harvard.edu/abs/2013MNRAS.435.1047B} {435, 1047}

\bibitem[\protect\citeauthoryear{{Chang}, {Wolf}, {Onken}, {Luvaul}, {Gupta}
  \& {Flynn}}{{Chang} et~al.}{2019a}]{2019ATel13008....1C}
{Chang} S.-W.,  {Wolf} C.,  {Onken} C.~A.,  {Luvaul} L.,  {Gupta} V.,   {Flynn}
  C.,  2019a, The Astronomer's Telegram, \href
  {https://ui.adsabs.harvard.edu/abs/2019ATel13008....1C} {13008, 1}

\bibitem[\protect\citeauthoryear{{Chang} et~al.,}{{Chang}
  et~al.}{2019b}]{Chang2019GCN.24260....1C}
{Chang} S.~W.,  et~al., 2019b, GRB Coordinates Network, \href
  {https://ui.adsabs.harvard.edu/abs/2019GCN.24260....1C} {24260, 1}

\bibitem[\protect\citeauthoryear{{Chang}, {Wolf}, {Onken}, {Luvaul}  \&
  {Scott}}{{Chang} et~al.}{2019c}]{Chang2019GCN.24325....1C}
{Chang} S.~W.,  {Wolf} C.,  {Onken} C.~A.,  {Luvaul} L.,   {Scott} S.,  2019c,
  GRB Coordinates Network, \href
  {https://ui.adsabs.harvard.edu/abs/2019GCN.24325....1C} {24325, 1}

\bibitem[\protect\citeauthoryear{{Chang}, {Wolf}  \& {Onken}}{{Chang}
  et~al.}{2020}]{Chang2020MNRAS.491...39C}
{Chang} S.-W.,  {Wolf} C.,   {Onken} C.~A.,  2020, \mndoi [\mnras]
  {10.1093/mnras/stz2898}, \href
  {https://ui.adsabs.harvard.edu/abs/2020MNRAS.491...39C} {491, 39}

\bibitem[\protect\citeauthoryear{Chen \& Guestrin}{Chen \&
  Guestrin}{2016}]{Chen10.1145/2939672.2939785}
Chen T.,  Guestrin C.,  2016, in Proceedings of the 22nd ACM SIGKDD
  International Conference on Knowledge Discovery and Data Mining. KDD ’16.
Association for Computing Machinery, New York, NY, USA, p. 785–794,
  \mndoi{10.1145/2939672.2939785}, \url
  {https://doi.org/10.1145/2939672.2939785}

\bibitem[\protect\citeauthoryear{{Coughlin} et~al.,}{{Coughlin}
  et~al.}{2019}]{2019ApJ...885L..19C}
{Coughlin} M.~W.,  et~al., 2019, \mndoi [\apjl] {10.3847/2041-8213/ab4ad8},
  \href {https://ui.adsabs.harvard.edu/abs/2019ApJ...885L..19C} {885, L19}

\bibitem[\protect\citeauthoryear{{Coulter} et~al.,}{{Coulter}
  et~al.}{2017}]{Coulter2017Sci...358.1556C}
{Coulter} D.~A.,  et~al., 2017, \mndoi [Science] {10.1126/science.aap9811},
  \href {https://ui.adsabs.harvard.edu/abs/2017Sci...358.1556C} {358, 1556}

\bibitem[\protect\citeauthoryear{{Cowperthwaite} et~al.,}{{Cowperthwaite}
  et~al.}{2017}]{Cowperthwaite2017ApJ...848L..17C}
{Cowperthwaite} P.~S.,  et~al., 2017, \mndoi [\apjl]
  {10.3847/2041-8213/aa8fc7}, \href
  {https://ui.adsabs.harvard.edu/abs/2017ApJ...848L..17C} {848, L17}

\bibitem[\protect\citeauthoryear{{D{\'\i}az} et~al.,}{{D{\'\i}az}
  et~al.}{2017}]{Diaz2017ApJ...848L..29D}
{D{\'\i}az} M.~C.,  et~al., 2017, \mndoi [\apjl] {10.3847/2041-8213/aa9060},
  \href {https://ui.adsabs.harvard.edu/abs/2017ApJ...848L..29D} {848, L29}

\bibitem[\protect\citeauthoryear{Dietterich}{Dietterich}{2000}]{Dietterich2000}
Dietterich T.~G.,  2000, Ensemble Methods in Machine Learning, 3 edn.
~MCS Vol. 1857, Springer-Verlag Berlin Heidelberg,
  https://doi.org/10.1007/3-540-45014-9\_1

\bibitem[\protect\citeauthoryear{{Dopita}, {Hart}, {McGregor}, {Oates},
  {Bloxham}  \& {Jones}}{{Dopita} et~al.}{2007}]{2007Ap&SS.310..255D}
{Dopita} M.,  {Hart} J.,  {McGregor} P.,  {Oates} P.,  {Bloxham} G.,   {Jones}
  D.,  2007, \mndoi [\apss] {10.1007/s10509-007-9510-z}, \href
  {https://ui.adsabs.harvard.edu/abs/2007Ap&SS.310..255D} {310, 255}

\bibitem[\protect\citeauthoryear{{Drout} et~al.,}{{Drout}
  et~al.}{2017}]{Drout2017Sci...358.1570D}
{Drout} M.~R.,  et~al., 2017, \mndoi [Science] {10.1126/science.aaq0049}, \href
  {https://ui.adsabs.harvard.edu/abs/2017Sci...358.1570D} {358, 1570}

\bibitem[\protect\citeauthoryear{{Duev} et~al.,}{{Duev}
  et~al.}{2019}]{Duev2019MNRAS.489.3582D}
{Duev} D.~A.,  et~al., 2019, \mndoi [\mnras] {10.1093/mnras/stz2357}, \href
  {https://ui.adsabs.harvard.edu/abs/2019MNRAS.489.3582D} {489, 3582}

\bibitem[\protect\citeauthoryear{{Farah} et~al.,}{{Farah}
  et~al.}{2018}]{2018MNRAS.478.1209F}
{Farah} W.,  et~al., 2018, \mndoi [\mnras] {10.1093/mnras/sty1122}, \href
  {https://ui.adsabs.harvard.edu/abs/2018MNRAS.478.1209F} {478, 1209}

\bibitem[\protect\citeauthoryear{{Farah} et~al.,}{{Farah}
  et~al.}{2019}]{2019MNRAS.488.2989F}
{Farah} W.,  et~al., 2019, \mndoi [\mnras] {10.1093/mnras/stz1748}, \href
  {https://ui.adsabs.harvard.edu/abs/2019MNRAS.488.2989F} {488, 2989}

\bibitem[\protect\citeauthoryear{{Fern{\'a}ndez} \& {Metzger}}{{Fern{\'a}ndez}
  \& {Metzger}}{2016}]{Fernandez2016ARNPS..66...23F}
{Fern{\'a}ndez} R.,  {Metzger} B.~D.,  2016, \mndoi [Annual Review of Nuclear
  and Particle Science] {10.1146/annurev-nucl-102115-044819}, \href
  {https://ui.adsabs.harvard.edu/abs/2016ARNPS..66...23F} {66, 23}

\bibitem[\protect\citeauthoryear{{Flesch}}{{Flesch}}{2015}]{Flesch2015PASA...32...10F}
{Flesch} E.~W.,  2015, \mndoi [\pasa] {10.1017/pasa.2015.10}, \href
  {https://ui.adsabs.harvard.edu/abs/2015PASA...32...10F} {32, e010}

\bibitem[\protect\citeauthoryear{Friedman}{Friedman}{2000}]{Friedman00greedyfunction}
Friedman J.~H.,  2000, Annals of Statistics, 29, 1189

\bibitem[\protect\citeauthoryear{{Gaia Collaboration} et~al.,}{{Gaia
  Collaboration} et~al.}{2016}]{2016A&A...595A...1G}
{Gaia Collaboration} et~al., 2016, \mndoi [\aap] {10.1051/0004-6361/201629272},
  \href {https://ui.adsabs.harvard.edu/abs/2016A&A...595A...1G} {595, A1}

\bibitem[\protect\citeauthoryear{{Gaia Collaboration}, {Brown}, {Vallenari},
  {Prusti}, {de Bruijne}, {Babusiaux}  \& {Biermann}}{{Gaia Collaboration}
  et~al.}{2020}]{2020arXiv201201533G}
{Gaia Collaboration} {Brown} A.~G.~A.,  {Vallenari} A.,  {Prusti} T.,  {de
  Bruijne} J.~H.~J.,  {Babusiaux} C.,   {Biermann} M.,  2020, arXiv e-prints,
  \href {https://ui.adsabs.harvard.edu/abs/2020arXiv201201533G} {p.
  arXiv:2012.01533}

\bibitem[\protect\citeauthoryear{{Goldstein} et~al.,}{{Goldstein}
  et~al.}{2015}]{Goldstein2015AJ....150...82G}
{Goldstein} D.~A.,  et~al., 2015, \mndoi [\aj] {10.1088/0004-6256/150/3/82},
  \href {https://ui.adsabs.harvard.edu/abs/2015AJ....150...82G} {150, 82}

\bibitem[\protect\citeauthoryear{{Guillochon}, {Parrent}, {Kelley}  \&
  {Margutti}}{{Guillochon} et~al.}{2017}]{Guillochon2017ApJ...835...64G}
{Guillochon} J.,  {Parrent} J.,  {Kelley} L.~Z.,   {Margutti} R.,  2017, \mndoi
  [\apj] {10.3847/1538-4357/835/1/64}, \href
  {https://ui.adsabs.harvard.edu/abs/2017ApJ...835...64G} {835, 64}

\bibitem[\protect\citeauthoryear{{Hjorth} et~al.,}{{Hjorth}
  et~al.}{2017}]{2017ApJ...848L..31H}
{Hjorth} J.,  et~al., 2017, \mndoi [\apjl] {10.3847/2041-8213/aa9110}, \href
  {https://ui.adsabs.harvard.edu/abs/2017ApJ...848L..31H} {848, L31}

\bibitem[\protect\citeauthoryear{{Hosseinzadeh} et~al.,}{{Hosseinzadeh}
  et~al.}{2019}]{2019ApJ...880L...4H}
{Hosseinzadeh} G.,  et~al., 2019, \mndoi [\apjl] {10.3847/2041-8213/ab271c},
  \href {https://ui.adsabs.harvard.edu/abs/2019ApJ...880L...4H} {880, L4}

\bibitem[\protect\citeauthoryear{{Hu} et~al.,}{{Hu}
  et~al.}{2017}]{2017SciBu..62.1433H}
{Hu} L.,  et~al., 2017, \mndoi [Science Bulletin] {10.1016/j.scib.2017.10.006},
  \href {https://ui.adsabs.harvard.edu/abs/2017SciBu..62.1433H} {62, 1433}

\bibitem[\protect\citeauthoryear{{Jedicke}}{{Jedicke}}{1996}]{1996AJ....111..970J}
{Jedicke} R.,  1996, \mndoi [\aj] {10.1086/117844}, \href
  {https://ui.adsabs.harvard.edu/abs/1996AJ....111..970J} {111, 970}

\bibitem[\protect\citeauthoryear{{Kasen}, {Metzger}, {Barnes}, {Quataert}  \&
  {Ramirez-Ruiz}}{{Kasen} et~al.}{2017}]{2017Natur.551...80K}
{Kasen} D.,  {Metzger} B.,  {Barnes} J.,  {Quataert} E.,   {Ramirez-Ruiz} E.,
  2017, \mndoi [\nat] {10.1038/nature24453}, \href
  {https://ui.adsabs.harvard.edu/abs/2017Natur.551...80K} {551, 80}

\bibitem[\protect\citeauthoryear{{Kasliwal} et~al.,}{{Kasliwal}
  et~al.}{2017}]{Kasliwal2017Sci...358.1559K}
{Kasliwal} M.~M.,  et~al., 2017, \mndoi [Science] {10.1126/science.aap9455},
  \href {https://ui.adsabs.harvard.edu/abs/2017Sci...358.1559K} {358, 1559}

\bibitem[\protect\citeauthoryear{{Kasliwal} et~al.,}{{Kasliwal}
  et~al.}{2019}]{2019MNRAS.tmpL..14K}
{Kasliwal} M.~M.,  et~al., 2019, \mndoi [\mnras] {10.1093/mnrasl/slz007}, \href
  {https://ui.adsabs.harvard.edu/abs/2019MNRAS.tmpL..14K} {}

\bibitem[\protect\citeauthoryear{{Kasliwal} et~al.,}{{Kasliwal}
  et~al.}{2020}]{2020ApJ...905..145K}
{Kasliwal} M.~M.,  et~al., 2020, \mndoi [\apj] {10.3847/1538-4357/abc335},
  \href {https://ui.adsabs.harvard.edu/abs/2020ApJ...905..145K} {905, 145}

\bibitem[\protect\citeauthoryear{{Killestein} et~al.,}{{Killestein}
  et~al.}{2021}]{2021arXiv210209892K}
{Killestein} T.~L.,  et~al., 2021, arXiv e-prints, \href
  {https://ui.adsabs.harvard.edu/abs/2021arXiv210209892K} {p. arXiv:2102.09892}

\bibitem[\protect\citeauthoryear{{Kilpatrick} et~al.,}{{Kilpatrick}
  et~al.}{2017}]{Kilpatrick2017Sci...358.1583K}
{Kilpatrick} C.~D.,  et~al., 2017, \mndoi [Science] {10.1126/science.aaq0073},
  \href {https://ui.adsabs.harvard.edu/abs/2017Sci...358.1583K} {358, 1583}

\bibitem[\protect\citeauthoryear{{Kruckow}, {Tauris}, {Langer}, {Kramer}  \&
  {Izzard}}{{Kruckow} et~al.}{2018}]{2018MNRAS.481.1908K}
{Kruckow} M.~U.,  {Tauris} T.~M.,  {Langer} N.,  {Kramer} M.,   {Izzard} R.~G.,
   2018, \mndoi [\mnras] {10.1093/mnras/sty2190}, \href
  {https://ui.adsabs.harvard.edu/abs/2018MNRAS.481.1908K} {481, 1908}

\bibitem[\protect\citeauthoryear{{LIGO Scientific Collaboration} \& {VIRGO
  Collaboration}}{{LIGO Scientific Collaboration} \& {VIRGO
  Collaboration}}{2019}]{2019GCN.24168....1L}
{LIGO Scientific Collaboration} {VIRGO Collaboration} 2019, GRB Coordinates
  Network, \href {https://ui.adsabs.harvard.edu/abs/2019GCN.24168....1L}
  {24168, 1}

\bibitem[\protect\citeauthoryear{{LIGO Scientific Collaboration} \& {Virgo
  Collaboration}}{{LIGO Scientific Collaboration} \& {Virgo
  Collaboration}}{2019}]{2019GCN.25606....1L}
{LIGO Scientific Collaboration} {Virgo Collaboration} 2019, GRB Coordinates
  Network, \href {https://ui.adsabs.harvard.edu/abs/2019GCN.25606....1L}
  {25606, 1}

\bibitem[\protect\citeauthoryear{{Levan} et~al.,}{{Levan}
  et~al.}{2017}]{Levan2017ApJ...848L..28L}
{Levan} A.~J.,  et~al., 2017, \mndoi [\apjl] {10.3847/2041-8213/aa905f}, \href
  {https://ui.adsabs.harvard.edu/abs/2017ApJ...848L..28L} {848, L28}

\bibitem[\protect\citeauthoryear{{Li} \& {Paczy{\'n}ski}}{{Li} \&
  {Paczy{\'n}ski}}{1998}]{Li1998ApJ...507L..59L}
{Li} L.-X.,  {Paczy{\'n}ski} B.,  1998, \mndoi [\apjl] {10.1086/311680}, \href
  {https://ui.adsabs.harvard.edu/abs/1998ApJ...507L..59L} {507, L59}

\bibitem[\protect\citeauthoryear{{Lipunov} et~al.,}{{Lipunov}
  et~al.}{2019}]{Lipunov2019GCN.24326....1L}
{Lipunov} V.,  et~al., 2019, GRB Coordinates Network, \href
  {https://ui.adsabs.harvard.edu/abs/2019GCN.24326....1L} {24326, 1}

\bibitem[\protect\citeauthoryear{{Luvaul}, {Onken}, {Wolf}, {Smillie}  \&
  {Sebo}}{{Luvaul} et~al.}{2017}]{Luvaul2017ASPC..512..393L}
{Luvaul} L.~C.,  {Onken} C.~A.,  {Wolf} C.,  {Smillie} J.~G.,   {Sebo} K.~M.,
  2017, in {Lorente} N.~P.~F.,  {Shortridge} K.,   {Wayth} R.,  eds,
  Astronomical Society of the Pacific Conference Series Vol. 512, Astronomical
  Data Analysis Software and Systems XXV. p.~393

\bibitem[\protect\citeauthoryear{{Metzger}}{{Metzger}}{2019}]{Metzger2019LRR....23....1M}
{Metzger} B.~D.,  2019, \mndoi [Living Reviews in Relativity]
  {10.1007/s41114-019-0024-0}, \href
  {https://ui.adsabs.harvard.edu/abs/2019LRR....23....1M} {23, 1}

\bibitem[\protect\citeauthoryear{{Metzger} et~al.,}{{Metzger}
  et~al.}{2010}]{Metzger2010MNRAS.406.2650M}
{Metzger} B.~D.,  et~al., 2010, \mndoi [\mnras]
  {10.1111/j.1365-2966.2010.16864.x}, \href
  {https://ui.adsabs.harvard.edu/abs/2010MNRAS.406.2650M} {406, 2650}

\bibitem[\protect\citeauthoryear{{Metzger}, {Bauswein}, {Goriely}  \&
  {Kasen}}{{Metzger} et~al.}{2015}]{Metzger2015MNRAS.446.1115M}
{Metzger} B.~D.,  {Bauswein} A.,  {Goriely} S.,   {Kasen} D.,  2015, \mndoi
  [\mnras] {10.1093/mnras/stu2225}, \href
  {https://ui.adsabs.harvard.edu/abs/2015MNRAS.446.1115M} {446, 1115}

\bibitem[\protect\citeauthoryear{{Metzger}, {Thompson}  \&
  {Quataert}}{{Metzger} et~al.}{2018}]{Metzger2018ApJ...856..101M}
{Metzger} B.~D.,  {Thompson} T.~A.,   {Quataert} E.,  2018, \mndoi [\apj]
  {10.3847/1538-4357/aab095}, \href
  {https://ui.adsabs.harvard.edu/abs/2018ApJ...856..101M} {856, 101}

\bibitem[\protect\citeauthoryear{{M{\"o}ller} et~al.,}{{M{\"o}ller}
  et~al.}{2019}]{2019IAUS..339....3M}
{M{\"o}ller} A.,  et~al., 2019, in {Griffin} R.~E.,  ed.,  International
  Astronomical Union Symposium Series Vol. 339, Southern Horizons in
  Time-Domain Astronomy. pp~3--6, \mndoi{10.1017/S1743921318002077}

\bibitem[\protect\citeauthoryear{{Mooley} et~al.,}{{Mooley}
  et~al.}{2018}]{2018Natur.554..207M}
{Mooley} K.~P.,  et~al., 2018, \mndoi [\nat] {10.1038/nature25452}, \href
  {https://ui.adsabs.harvard.edu/abs/2018Natur.554..207M} {554, 207}

\bibitem[\protect\citeauthoryear{Onken et~al.,}{Onken et~al.}{2019}]{Onken2019}
Onken C.~A.,  et~al., 2019, \mndoi [Publications of the Astronomical Society of
  Australia] {10.1017/pasa.2019.27}, 36, e033

\bibitem[\protect\citeauthoryear{{Petroff} et~al.,}{{Petroff}
  et~al.}{2015}]{2015MNRAS.447..246P}
{Petroff} E.,  et~al., 2015, \mndoi [\mnras] {10.1093/mnras/stu2419}, \href
  {https://ui.adsabs.harvard.edu/abs/2015MNRAS.447..246P} {447, 246}

\bibitem[\protect\citeauthoryear{{Pian} et~al.,}{{Pian}
  et~al.}{2017}]{Pian2017Natur.551...67P}
{Pian} E.,  et~al., 2017, \mndoi [\nat] {10.1038/nature24298}, \href
  {https://ui.adsabs.harvard.edu/abs/2017Natur.551...67P} {551, 67}

\bibitem[\protect\citeauthoryear{{Price} et~al.,}{{Price}
  et~al.}{2019}]{2019MNRAS.486.3636P}
{Price} D.~C.,  et~al., 2019, \mndoi [\mnras] {10.1093/mnras/stz958}, \href
  {https://ui.adsabs.harvard.edu/abs/2019MNRAS.486.3636P} {486, 3636}

\bibitem[\protect\citeauthoryear{{Romero-Shaw}, {Farrow}, {Stevenson}, {Thrane}
   \& {Zhu}}{{Romero-Shaw} et~al.}{2020}]{Romero-Shaw2020arXiv200106492R}
{Romero-Shaw} I.~M.,  {Farrow} N.,  {Stevenson} S.,  {Thrane} E.,   {Zhu}
  X.-J.,  2020, arXiv e-prints, \href
  {https://ui.adsabs.harvard.edu/abs/2020arXiv200106492R} {p. arXiv:2001.06492}

\bibitem[\protect\citeauthoryear{{Safarzadeh}, {Ramirez-Ruiz}  \&
  {Berger}}{{Safarzadeh} et~al.}{2020}]{Safarzadeh2020arXiv200104502S}
{Safarzadeh} M.,  {Ramirez-Ruiz} E.,   {Berger} E.,  2020, arXiv e-prints,
  \href {https://ui.adsabs.harvard.edu/abs/2020arXiv200104502S} {p.
  arXiv:2001.04502}

\bibitem[\protect\citeauthoryear{{Scalzo} et~al.,}{{Scalzo}
  et~al.}{2017}]{Scalzo2017PASA...34...30S}
{Scalzo} R.~A.,  et~al., 2017, \mndoi [\pasa] {10.1017/pasa.2017.24}, \href
  {https://ui.adsabs.harvard.edu/abs/2017PASA...34...30S} {34, e030}

\bibitem[\protect\citeauthoryear{{Shappee} et~al.,}{{Shappee}
  et~al.}{2017}]{Shappee2017Sci...358.1574S}
{Shappee} B.~J.,  et~al., 2017, \mndoi [Science] {10.1126/science.aaq0186},
  \href {https://ui.adsabs.harvard.edu/abs/2017Sci...358.1574S} {358, 1574}

\bibitem[\protect\citeauthoryear{Singer \& Price}{Singer \&
  Price}{2016}]{PhysRevD.93.024013}
Singer L.~P.,  Price L.~R.,  2016, \mndoi [Phys. Rev. D]
  {10.1103/PhysRevD.93.024013}, 93, 024013

\bibitem[\protect\citeauthoryear{{Singer} et~al.,}{{Singer}
  et~al.}{2016}]{2016ApJS..226...10S}
{Singer} L.~P.,  et~al., 2016, \mndoi [\apjs] {10.3847/0067-0049/226/1/10},
  \href {https://ui.adsabs.harvard.edu/abs/2016ApJS..226...10S} {226, 10}

\bibitem[\protect\citeauthoryear{{Smartt} et~al.,}{{Smartt}
  et~al.}{2017}]{Smartt2017Natur.551...75S}
{Smartt} S.~J.,  et~al., 2017, \mndoi [\nat] {10.1038/nature24303}, \href
  {https://ui.adsabs.harvard.edu/abs/2017Natur.551...75S} {551, 75}

\bibitem[\protect\citeauthoryear{{Tanaka} \& {Hotokezaka}}{{Tanaka} \&
  {Hotokezaka}}{2013}]{Tanaka2013ApJ...775..113T}
{Tanaka} M.,  {Hotokezaka} K.,  2013, \mndoi [\apj]
  {10.1088/0004-637X/775/2/113}, \href
  {https://ui.adsabs.harvard.edu/abs/2013ApJ...775..113T} {775, 113}

\bibitem[\protect\citeauthoryear{{Tanvir} et~al.,}{{Tanvir}
  et~al.}{2017}]{Tanvir2017ApJ...848L..27T}
{Tanvir} N.~R.,  et~al., 2017, \mndoi [\apjl] {10.3847/2041-8213/aa90b6}, \href
  {https://ui.adsabs.harvard.edu/abs/2017ApJ...848L..27T} {848, L27}

\bibitem[\protect\citeauthoryear{{Troja} et~al.,}{{Troja}
  et~al.}{2017}]{2017Natur.551...71T}
{Troja} E.,  et~al., 2017, \mndoi [\nat] {10.1038/nature24290}, \href
  {https://ui.adsabs.harvard.edu/abs/2017Natur.551...71T} {551, 71}

\bibitem[\protect\citeauthoryear{{Troja} et~al.,}{{Troja}
  et~al.}{2018}]{2018NatCo...9.4089T}
{Troja} E.,  et~al., 2018, \mndoi [Nature Communications]
  {10.1038/s41467-018-06558-7}, \href
  {https://ui.adsabs.harvard.edu/abs/2018NatCo...9.4089T} {9, 4089}

\bibitem[\protect\citeauthoryear{{Utsumi} et~al.,}{{Utsumi}
  et~al.}{2017}]{Utsumi2017PASJ...69..101U}
{Utsumi} Y.,  et~al., 2017, \mndoi [\pasj] {10.1093/pasj/psx118}, \href
  {https://ui.adsabs.harvard.edu/abs/2017PASJ...69..101U} {69, 101}

\bibitem[\protect\citeauthoryear{Veitch et~al.,}{Veitch
  et~al.}{2015}]{PhysRevD.91.042003}
Veitch J.,  et~al., 2015, \mndoi [Phys. Rev. D] {10.1103/PhysRevD.91.042003},
  91, 042003

\bibitem[\protect\citeauthoryear{{Watson}, {Henden}  \& {Price}}{{Watson}
  et~al.}{2006}]{Watson2006SASS...25...47W}
{Watson} C.~L.,  {Henden} A.~A.,   {Price} A.,  2006, Society for Astronomical
  Sciences Annual Symposium, \href
  {https://ui.adsabs.harvard.edu/abs/2006SASS...25...47W} {25, 47}

\bibitem[\protect\citeauthoryear{{Wolf} et~al.,}{{Wolf}
  et~al.}{2018}]{Wolf2018PASA...35...10W}
{Wolf} C.,  et~al., 2018, \mndoi [\pasa] {10.1017/pasa.2018.5}, \href
  {https://ui.adsabs.harvard.edu/abs/2018PASA...35...10W} {35, e010}

\bibitem[\protect\citeauthoryear{{Wright} et~al.,}{{Wright}
  et~al.}{2015}]{Wright2015MNRAS.449..451W}
{Wright} D.~E.,  et~al., 2015, \mndoi [\mnras] {10.1093/mnras/stv292}, \href
  {https://ui.adsabs.harvard.edu/abs/2015MNRAS.449..451W} {449, 451}

\bibitem[\protect\citeauthoryear{{Wu}, {Barnes}, {Mart{\'\i}nez-Pinedo}  \&
  {Metzger}}{{Wu} et~al.}{2019}]{2019PhRvL.122f2701W}
{Wu} M.-R.,  {Barnes} J.,  {Mart{\'\i}nez-Pinedo} G.,   {Metzger} B.~D.,  2019,
  \mndoi [\prl] {10.1103/PhysRevLett.122.062701}, \href
  {https://ui.adsabs.harvard.edu/abs/2019PhRvL.122f2701W} {122, 062701}

\bibitem[\protect\citeauthoryear{{van Roestel} et~al.,}{{van Roestel}
  et~al.}{2019}]{vanRoestel2019MNRAS.484.4507V}
{van Roestel} J.,  et~al., 2019, \mndoi [\mnras] {10.1093/mnras/stz241}, \href
  {https://ui.adsabs.harvard.edu/abs/2019MNRAS.484.4507V} {484, 4507}

\makeatother
\end{thebibliography}

\end{document}